\documentclass[preprint,12pt]{elsarticle}
\usepackage{multirow}
\usepackage[utf8]{inputenc}
\usepackage{amssymb}
\journal{Annals of Physics}

\begin{document}

\begin{frontmatter}

%% Title, authors and addresses

%% use the tnoteref command within \title for footnotes;
%% use the tnotetext command for theassociated footnote;
%% use the fnref command within \author or \affiliation for footnotes;
%% use the fntext command for theassociated footnote;
%% use the corref command within \author for corresponding author footnotes;
%% use the cortext command for theassociated footnote;
%% use the ead command for the email address,
%% and the form \ead[url] for the home page:
%% \title{Title\tnoteref{label1}}
%% \tnotetext[label1]{}
%% \author{Name\corref{cor1}\fnref{label2}}
%% \ead{email address}
%% \ead[url]{home page}
%% \fntext[label2]{}
%% \cortext[cor1]{}
%% \affiliation{organization={},
%%            addressline={},
%%            city={},
%%            postcode={},
%%            state={},
%%            country={}}
%% \fntext[label3]{}

\title{Random Cantor sets and mini-bands in local spectrum of quantum systems}

%% use optional labels to link authors explicitly to addresses:
%% \author[label1,label2]{}
%% \affiliation[label1]{organization={},
%%             addressline={},
%%             city={},
%%             postcode={},
%%             state={},
%%             country={}}
%%
%% \affiliation[label2]{organization={},
%%             addressline={},
%%             city={},
%%             postcode={},
%%             state={},
%%             country={}}
\author[1]{B. L. Altshuler}
\author[2]{V. E. Kravtsov}
%\address[1]{Department of Physics, Columbia University, 116th and Broadway, New York, NY, 10027, U.S.A.} 
%\address[2]{The Abdus Salam International Centre for Theoretical Physics, P.O.B. 586, Trieste, 34100, Italy}
\affiliation[1]{organization={Department of Physics, Columbia University},
            addressline={116th and Broadway},
            city={New York},
            postcode={10027},
            state={NY},
            country={U.S.A.}}
\affiliation[2]{organization={The Abdus Salam International Centre for
Theoretical Physics},%Department and Organization
            addressline={P.O.B. 586},
            city={Trieste},
            postcode={34100},
            country={Italy}}

\begin{abstract}
In this paper we give a physically transparent picture of singular-continuous spectrum in disordered systems which possess a non-ergodic extended phase. We present a simple model of identically and independently distributed level spacing in the spectrum of local density of states
and show how a fat tail appears in this distribution at the broad distribution of eigenfunction amplitudes. For the model with a power-law local spacing distribution we derive the correlation function $K(\omega)$ of the local density of states and show that depending on the relation between the eigenfunction fractal dimension $D_{2}$ and the spectral fractal dimension $D_{s}$ encoded in the power-law spacing distribution, a singular continuous spectrum of a random Cantor set or that of an isolated mini-band may appear. In the limit of an infinite number of degrees of freedom the function $K(\omega)$ in the non-ergodic extended phase is singular at $\omega=0$ with the branch-cut singularity for the case of a random Cantor set and with the $\delta$-function singularity for the case of an isolated mini-band. For an absolutely continuous spectrum $K(\omega)$ tends to a finite limit as $\omega\rightarrow 0$. For an arbitrary local spacing distribution function we formulated a criterion of fractality of local spectrum and tested it on simple examples.

\end{abstract}

%%Graphical abstract
\begin{graphicalabstract}
\includegraphics[width=0.5\linewidth]{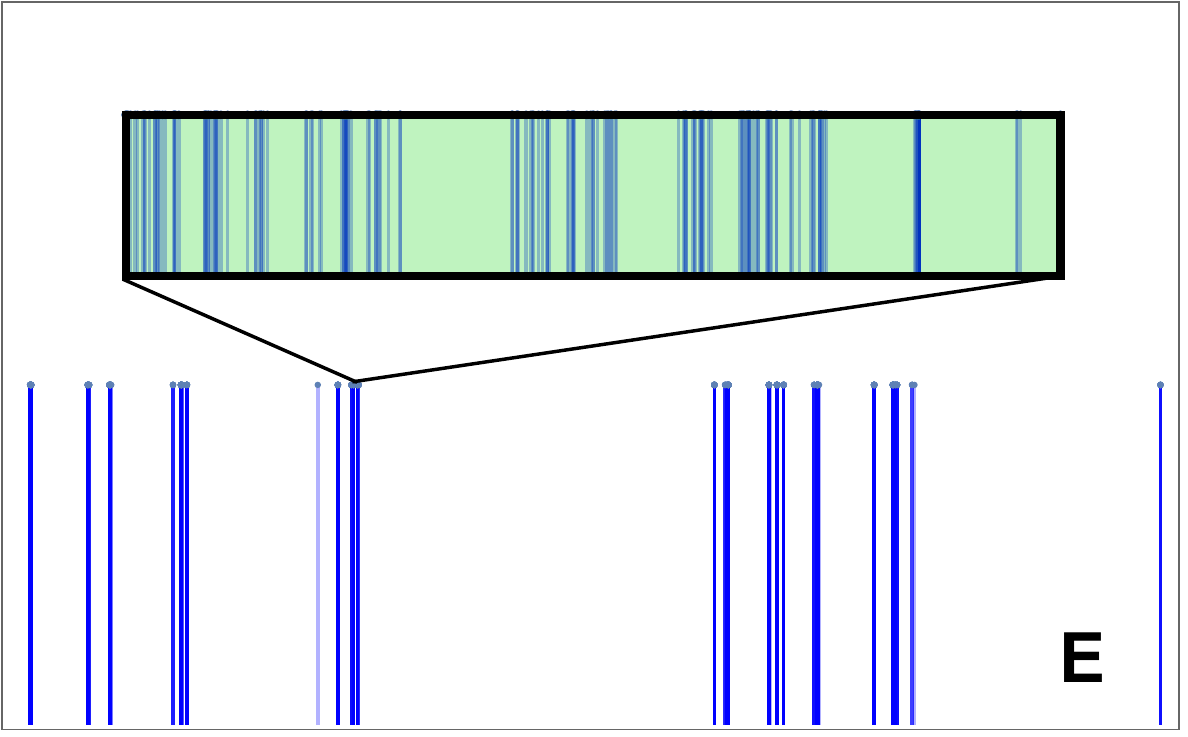}
\end{graphicalabstract}

%%Research highlights
\begin{highlights}
\item Non-ergodic extended states  and a singular-continuous (s.c.)  spectrum
\item Two types of s.c. spectrum:   Cantor set and isolated mini-band
\item Model for s.c. spectrum:  i.i.d. power-law spacing distribution
\item  Correlation function of LDoS $K(\omega)$ as a measure for s.c. spectrum
\item Singularity of $K(\omega)$ at $\omega=0$  for s.c. spectrum in the infinite system
\end{highlights}

\begin{keyword}
 non-ergodic extended states, multifractality of wave functions, singular-continuous spectrum, Cantor set, mini-bands

%% PACS codes here, in the form: \PACS code \sep code

%% MSC codes here, in the form: \MSC code \sep code
%% or \MSC[2008] code \sep code (2000 is the default)

\end{keyword}

\end{frontmatter}

%% \linenumbers

%% main text
\section{Preamble}
This paper is dedicated to memory of our friend, collaborator and teacher Kostya Efetov. His scientific fearlessness, persistent innovation and  human qualities made an unforgettable and crucial impact on our scientific careers, as well as on the development of our scientific community as a whole.
\section{Introduction}
\label{Int}
Recently there has been  a vigorous discussion
on the existence of {\it non-ergodic extended (NEE) states}  in a finite range of parameters (NEE {\it phase}) of various disordered systems, both free-particle \cite{Biroli_Tarzia_unpub, DeLuca2014,RP15,FacoetiBiroli_RP,Warzel, AnnalRRG,LN-RP20,Lemarie2017,TikhMir19, beta} and interacting \cite{Pino-Ioffe-VEK,Laflorencie19,luitz2019multifractality,Baecker2019,TikhMir_toy, KhayPollWarWarzel, Khay_Gorsk,DeLuca_Scard,Tarzia_MBL_Hilbert}. By {\it extended} states one denotes the states that occupy an extensive part of the Hilbert space, while the lack of ergodicity implies that not all the Hilbert space is occupied by a wave function. The most well-known example of such  NEE states are the multifractal states at the critical point of the Anderson localization transition \cite{Mirlin-rev}. They, however, do not form a {\it phase}. Another example of such states are the states in the Critical Power-Law Banded Random Matrices \cite{PLBRM} which can be viewed as a one-dimensional systems with long-range hopping. In this case, however, all the states have an NEE, multifractal character. The first and the simplest system which possesses the NEE phase separated by the {\it ergodic transition} from the ergodic states and by the Anderson {\it localization transition} from the localized states was found   in  the generalized Gaussian Rosenzweig-Porter (RP) model  \cite{RP60, RP15} and further elaborated in
\cite{FacoetiBiroli_RP,LN-RP20, Monthus17b} to finally bring it to the status of a mathematical theorem \cite{Warzel}.

Remarkably, in Ref.\cite{FacoetiBiroli_RP}  the question of the structure of the local spectrum in the Rosenzweig-Porter model was raised and it was shown that this spectrum has a non-trivial form of a {\it mini-band}. This peculiarity of the spectrum determines the local dynamics of this model, e.g. the return, or survival, probability, which was studied in detail in Ref.\cite{return}. Furthermore, a modification of the RP model, the log-normal RP model \cite{LN-RP20},  was shown \cite{LN-RP-RRG21} to exhibit a {\it sub-difusive}, stretch-exponential relaxation of the averaged survival probability, similar to that in the extended phase of the Anderson localization model on a Random Regular Graph \cite{RRG_R(t),deTomasi2019subdiffusion} and in interacting spin-chains \cite{BarLev15,anom_diff,Luitz_Laflorencie_Alet, luitz2016,Scardicchio_subdiff_2018, Bardarson19,luitz2019multifractality}. Since the character of local spectrum determines local
dynamics and the latter is important for manipulation of the chains of qubits (which are equivalent to interacting spin-chains), the problem of local spectra has a wide application in quantum computing.

There is, however, another, fundamental dimension in this problem. From the mathematical literature \cite{Dunford} it is known that the spectrum of any Hamiltonian can be decomposed into a sum of three distinctly different types of spectra: the
pure-point, absolutely continuous and singular-continuous.
Their  loose definitions are the following:
\begin{itemize}
\item
{\bf $(a)$ pure point:} This type of spectrum corresponds to $L^{2}$ integrable {\it eigenfunctions}, e.g.   the {\it localized} wave functions with a finite localization radius. Finite-amplitude delocalized wave functions (e.g. the plane waves) and those localized in one single point [e.g. $\psi\sim \delta(p-p_{0})$ in the momentum space] are not $L^{2}$-integrable in an infinite  continuous system and, therefore, they cannot correspond to a pure point spectrum according to the mathematical literature.  By {\it spectrum} we understand the set of real numbers $\{E_{n}\}$ such that the operator $\hat{H}-E_{n}$ is not invertible.
\item
{\bf $(b)$ absolutely-continuous:} In this case the wave functions are   {\it not} $L^{2}$ integrable (not normalizable). These maybe  the delocalized wave functions in an infinite coordinate space, or the corresponding non-normalizable in the continuous space functions divergent strongly enough at some point $p$ in the momentum space. For this type of spectrum the set of $E_{n}$ must be {\it dense}, i.e. for any given real $E$ inside the spectrum  there is a $E_{n}$ such that $|E-E_{n}|$ is arbitrary small.
\item
{\bf $(c)$ singular-continuous:} Loosely speaking, this is the type of spectrum that does not fall in neither of the above two types $(a)$ and $(b)$. This means that the corresponding wave functions are {\it extended} in some basis (otherwise they belong to $(a)$) but the set of $E_{n}$ is {\it not dense}. The archetypical example of such a spectrum is the regular Cantor set. Indeed, in the Cantor set there is a hierarchy of gaps (nearest neighbor spacings) $\delta_{m}=3^{-m}$ ($m=1,2,...,\ln N/\ln3$) which values in the limit $1/N\rightarrow 0$ are accumulated  near $\delta_{\infty}=0$. Therefore  any point $E$ inside such a spectrum   falls inside some gap $\delta_{n}$ with a fixed {\it finite} $n$ which does not shrink to zero as $1/N\rightarrow 0$. Instead, with increasing $N$ more small gaps of the higher generations $m\gg n$ appear and accumulate near $\delta_{\infty}=0$. Hence, the distance $|E-E_{n}|$ with $E\in \delta_{n}$ remains finite in this limit, and the spectrum is {\it nowhere dense} albeit contains an infinite number $N\rightarrow \infty$ of levels.
\end{itemize}
 The above definitions make sense in the continuous systems, while the most of physical systems are defined on discrete sites of some lattices or graphs. In such systems at an infinite number of sites $N$ the non-normalizability can happen only for extended wave functions, while any localized wave function is always normalizable. However, the Fourier transform of the extended wave function is a localized one and vice versa.  Thus  the normalizability in the infinite discrete systems is not basis-invariant, while the global spectrum is. This means that normalizability in   discrete systems cannot be used for classification of basis-invariant global spectra  like in the above definitions for the continuous systems.

However, even in discrete systems one may classify the spectra but in a non-invariant way associated with a certain basis. In order to accomplish this goal we  consider the {\it local density of states} (LDoS) and its correlation functions where the spectral $\delta$-functions $\delta(E-E_{n})$ are weighted by the amplitudes $|\psi_{n}(r)|^{2}$ of the {\it normalized} eigenfunctions at a position $r$ in a chosen basis. However, if one adopts this rout   one should inevitably consider a finite dimension of the Hilbert space $N$, as otherwise the amplitude of an {\it extended} eigenfunction will be zero.  In such a discrete finite-$N$ systems the spectrum is always discrete and the solutions $\psi_{n}(r)$ of the Schroedinger equation are always normalizable. The limit $N\rightarrow\infty$ should be implemented only in the final results for the correlation functions of the LDoS.

The question is: Which are the analogues of the pure point, absolutely continuous and singular continuous for the {\it local} spectra (e.g. the spectra seen in LDoS) and how to distinguish statistically between them?  The simple solution we propose is to study the behavior of the local density of states correlation function $K(\omega)$ in the limit $N\rightarrow\infty$, where $\omega$ is the difference in the energy. We show that for a local analogue of the absolutely continuous spectrum (referred to as $l$-absolutely continuous spectrum) this correlation function has a finite limit at $\omega\rightarrow 0$, while for the $l$-singular continuous spectrum $K(\omega)$ is singular at $\omega=0$. For a $l$-pure point spectrum $K(\omega\rightarrow 0)=0$,  the limit $N\rightarrow \infty$ should be taken first in all three cases.

A crucial step in doing this limit is to discriminate eigenstates according to their amplitude at the observation point $r$ and eliminate the states which amplitude $|\psi_{i}(r)|^{2}$ is lower than the discriminant level $\sim 1/N$.
In the non-ergodic phases, this operation diminishes drastically the number of states remaining in the spectrum of LDoS (the "local levels") and it also changes qualitatively statistics of spacing between the corresponding energy levels leading to the fat tails in the level spacing distribution.

We propose the following classification of the local spectra:
\begin{itemize}
\item {\it $l$-pure point spectrum} $(a)$: This type of local  spectrum emerges in the localized systems. The local spectrum consists of a finite number of levels with the typical spacing that remains finite as $N\rightarrow\infty$. The corresponding states are all localized in the same localization volume.
\item {\it $l$-absolutely continuous spectrum} $(b)$: The local spectrum is dense in the limit $N\rightarrow\infty$ in a certain spectral interval that remains finite in this limit.   This type of spectrum corresponds to the {\it ergodic delocalized states}.
\item {\it $l$-singular continuous spectrum} $(c)$: These are the local spectra that do not fall in neither of the above types (a) and (b). Two important examples are:  ${\bf (i)}$ The local spectrum contains an infinite number of levels in a certain finite interval of energies in the limit $N\rightarrow\infty$. However, the spectrum is not dense in this interval (e.g. it is a random Cantor set). Alternatively, ${\bf (ii)}$ the local spectrum is all concentrated within a {\it mini-band} that contains an infinite number of levels in the $N\rightarrow\infty$ limit but the width of the mini-band shrinks to zero in this limit.

We will show below that in the case ${\bf (i)}$ $K(\omega)\sim \omega^{-\alpha}$, $0<\alpha<1$ in the limit $N\rightarrow\infty$, while in the case ${\bf (ii)}$ $K(\omega)\sim \delta(\omega)$ in this limit.
\end{itemize}

In order to substantiate this classification we consider a simple model of local levels with independent identically distributed (i.i.d.) spacing which distribution is a power-law. We show that such a simple model is capable of reproducing the two principle cases of $l$-singular continuous spectra: the random Cantor set and the isolated mini-band. We show that upon a proper choice of parameters such a model gives a correct   behavior of $K(\omega)$ both for the Power-Law Banded Random Matrices   and for the Gaussian Rosenzweig-Porter random matrix ensemble.  This allows us to conclude that the local spectrum of the former is a random Cantor set, while that of the latter is an isolated mini-band with the dense spectrum inside a mini-band.

However, our model gives many more possibilities of singular continuous spectra which Hamiltonian realizations are not yet found. Thus it can be viewed as a guide for a further search of such models.

Finally, we extend our model of i.i.d. local spacings to an {it arbitrary} distribution of spacings and formulate a general criterion of fractality of local spectrum. This criterion may be used to characterize complex statistics of local spectra in random systems.

The paper is organized as follows. In Introduction (Section 2) we review the relevant literature and discuss the difference between the classification of global spectra in continuous systems in mathematical literature and classification of local spectra in descrete systems in this work. In Section 3 we describe the procedure of discrimination of eigenstates which explains a qualitative difference between the global and the local spectral statistics. The two principle classes of the $l$-singular continuous spectra and the corresponding Hamiltonian realizations are also described in this section. In Section 4 a toy model of local spectrum of the form of  the textbook Cantor set is considered, the power-law spacing distribution $P(s)$ in this model is derived   and the model  local density of states correlation function $K(\omega)$ is computed. In Section 5 a random Cantor set is constructed from the i.i.d. power-law local spacing distributions and the correlation function $K(\omega)$ is computed. In Section 6  from the same model of i.i.d. local spacing distributions we obtain an   isolated mini-band. In Section 7   a general criterion of fractality of local spectrum is formulated for an arbitrary $P(s)$. This formalism is tested in Appendix A  for a power-law $P(s)$ for a choice of parameters that corresponds to a random Cantor set and  to an isolated mini-band in the NEE phases and to a dense spectrum in ergodic phases. In Conclusion we present the table of all the principle results for the $\omega$- and $N$- dependence of the local DoS correlation function $K(\omega)$ and for the scaling of the mini-band with the dimension $N$ of the Hilbert space. Finally, in Conclusion we present an expression for $K(\omega)$ in $N\rightarrow\infty$ limit which distinguishes between the absolutely continuous and the two principal classes of the singular continuous spectra that emerge in our model.

%%%%%%%%%%%%%%%%%%%%%%%%%%%%%%%%%%%%%%%%%%%%%%%%%%%%%%%%%%%%%%%%%%%%%%%%%%%%%%%%%%%%%%%%%%%%%%%%%%%%%%%%%%%%%%%%%%%
\section{Discrimination of eigenstates}\label{discr}
%%%%%%%%%%%%%%%%%%%%%%%%%%%%%%%%%%%%%%%%%%%%%%%%%%%%%%%%%%%%%%%%%%%%%%%%%%%%%%%%%%%%%%%%%%%%%%%%%%%%%%%%%%%%%%%%%%%
\subsection{The global and the local density of states}
We start by reviewing the differences between the global an the local spectrum of a quantum system with the broad distribution of the eigenfunction amplitudes $|\psi_{n}(r)|^{2}$. In this section we introduce the notion of a mini-band in the local spectrum and the peculiarities of the local spacing distribution compared to its global counterpart.

First of all we remind the readers the definitions of the global, $\rho(E)$, and the local, $\rho(E;r)$, densities of states (DoS) in terms of the eigenvalues $E_{n}$ and eigenfunctions $\psi_{n}(r)$ :
\begin{equation}\label{global_DoS}
\rho(E)=\frac{1}{N}\sum_{n}\delta(E-E_{n}),
\end{equation}
\begin{equation}\label{local_DoS}
\rho(E;r)=\sum_{n}|\psi_{n}(r)|^{2}\,\delta(E-E_{n}).
\end{equation}
In disordered systems both $\rho(E)$ and $\rho(E;r)$ are random quantities. By normalization of the eigenfunctions:
\begin{equation}\label{norm}
\sum_{r}|\psi_{n}(r)|^{2}=1,
\end{equation}
the mean eigenfunction amplitude is given by:
\begin{equation}
\langle |\psi_{n}(r)|^{2}\rangle=N^{-1},
\end{equation}
where $\langle ...\rangle$ denotes the ensemble average and $N$ is the dimension of the Hilbert space (the total number of eigenfunctions).

If the distribution of the eigenfunction amplitudes is narrow $|\psi_{n}(r)|\sim N^{-1}$ (as it happens e.g. for the Wigner-Dyson random matrices) then there is no qualitative difference between the statistics of the global and the local DoS. On the other hand, for the localized eigenfunctions only few of them have an appreciable amplitude in the observation point $r$. If one discards the states with exponentially small amplitudes in the observation point, the local spectrum will consist of a finite number of the $\delta$-functions with  a finite separations between them in the thermodynamic limit $N\rightarrow\infty$ (an $l$-pure point spectrum). In contrast, the global spectrum in most of the cases has a dense set of eigenvalues with the mean level spacing $\delta\sim N^{-1}$ that vanishes in the limit $N\rightarrow\infty$. In what follows we will always assume that the global spectrum is {\it dense}, i.e. in the thermodynamic limit  in the arbitrary close vicinity $\epsilon$ of any point $E$ inside the spectral band there is an eigenvalue $E_{n}$ such that $|E-E_{n}|<\epsilon$.

The two limiting examples mentioned above: (i) the absolutely continuous local and global spectra of the Wigner-Dyson random matrices \cite{Mehta} and (ii) the pure point local spectrum concomitant with absolutely continuous global spectrum of localized eigenstates, do not cover all the possible cases.
%%%%%%%%%%%%%%%%%%%%%%%%%%%%%%%%%%%%%%%%%%%%%%%%%%%%%%%%%%%%%%%%%%%%%%%%%%%%%%%%%%%%%%%%%%%%%%%%%%%%%%%%%%%%%%%%%%%%%%%%%%%%%%%%%%%%%%%%%%%%%
\begin{figure}[t]
\center{
\includegraphics[width=0.7\linewidth]{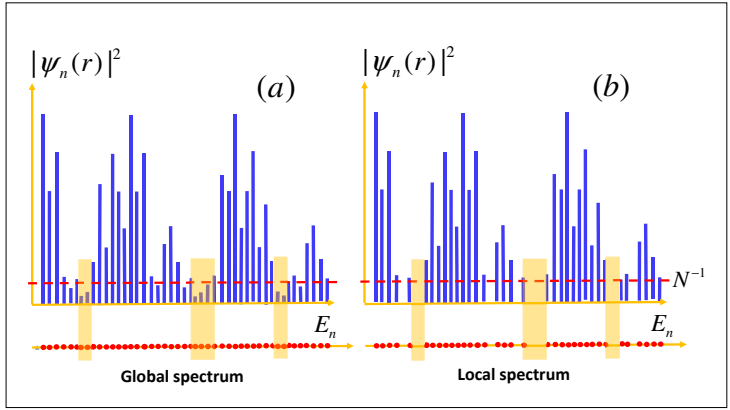}
}
\caption{(Color online) Eigenstates before (a) and after (b) discarding states with the amplitude at the observation point $r$ smaller than the discrimination level $\sim N^{-1}$ (red dashed lines). As the result of discrimination large gaps appear in the local spectrum (e.g. in the yellow shadowed areas). }
\label{Fig:large_gaps}
\end{figure}	
%%%%%%%%%%%%%%%%%%%%%%%%%%%%%%%%%%%%%%%%%%%%%%%%%%%%%%%%%%%%%%%%%%%%%%%%%%%%%%%%%%%%%%%%%%%%%%%%%%%%%%%%%%%%%%%%%%%%%%%%%%%%%%%%%%%%%%%%%%%%%
Below we describe other possibilities of the {\it $l$-singular continuous spectra} and present a transparent physical picture as to why and where they may arise.

 %%%%%%%%%%%%%%%%%%%%%%%%%%%%%%%%%%%%%%%%%%%%%%%%%%%%%%%%%%%%%%%%%%%%%%%%%%%%%%%%%%%%%%%%%%%%%%%%%%%%%%%%%%%%%%%%%%%
   \subsection{Discrimination of eigenstates and the fat tail in the local level spacing distribution}
%%%%%%%%%%%%%%%%%%%%%%%%%%%%%%%%%%%%%%%%%%%%%%%%%%%%%%%%%%%%%%%%%%%%%%%%%%%%%%%%%%%%%%%%%%%%%%%%%%%%%%%%%%%%%%%%%%%

 The main cause for a drastic difference between the statistics of the local and the global spectrum is best illustrated by the example of the localized eigenstates. It is the {\it broad distribution} of the eigenfunction amplitudes. In the case of localized eigenstates there are  rare large amplitudes if the localization center is close to the observation point and the range of exponentially small amplitudes if the observation point is beyond the localization volume. This allows to introduce a {\it discriminant} such that the eigenstates with amplitudes smaller than the discriminant level are discarded from the local spectrum (see Fig.\ref{Fig:large_gaps}). In practice the discriminant level should be chosen to be much lower than the mean eigenfunction amplitude
$\langle |\psi_{n}(r)|^{2}\rangle=N^{-1}$. The obvious consequence of that is the emergence in the local spectrum of level spacings much larger than those in the global one (see Fig.\ref{Fig:large_gaps}). Statistically, this exhibits itself in the fat tails in the local spacing distribution function.

An important special case is when {\it after the discrimination} the number of states in the local spectrum $M$ is infinite in the thermodynamic limit but it scales with $N$ like a fractional power of $N$:
\begin{equation}\label{D_s}
M\sim N^{D_{s}},\;\;\;0<D_{s}<1
\end{equation}
In this case the mean level spacing $\delta_{{\rm av}}$ in the local spectrum scales like $M^{-1}\sim N^{-D_{s}}$, while the typical level spacing $\delta_{{\rm typ}}\sim \delta$ is supposed to be the same  as in the global one \footnote{In general, the discrimination procedure insures only that $\delta_{{\rm typ}}\gtrsim \delta$. However, it seems natural that for the extended states the amplitudes are correlated at a scale of $\delta$. Therefore if a state survives the discrimination, the neighboring in energy state is likely to do so.}:
\begin{equation}\label{delta_typ-av}
\delta_{{\rm typ}}\approx \delta= 1/N,\;\;\;\delta_{{\rm av}}\sim N^{-D_{s}}.
\end{equation}
 	
Here and everywhere throughout the paper we assume that the energy units are chosen so that the global spectral band-width is equal to 1.
%%%%%%%%%%%%%%%%%%%%%%%%%%%%%%%%%%%%%%%%%%%%%%%%%%%%%%%%%%%%%%%%%%%%%%%%%%%%%%%%%%%%%%%%%%%%%%%%%%%%%%%%%%%%%%%%%%%%%%%%%%%%%%%%%%%%%%%%%%
\begin{figure}[tbh]
\center{
\includegraphics[width=0.45\linewidth]{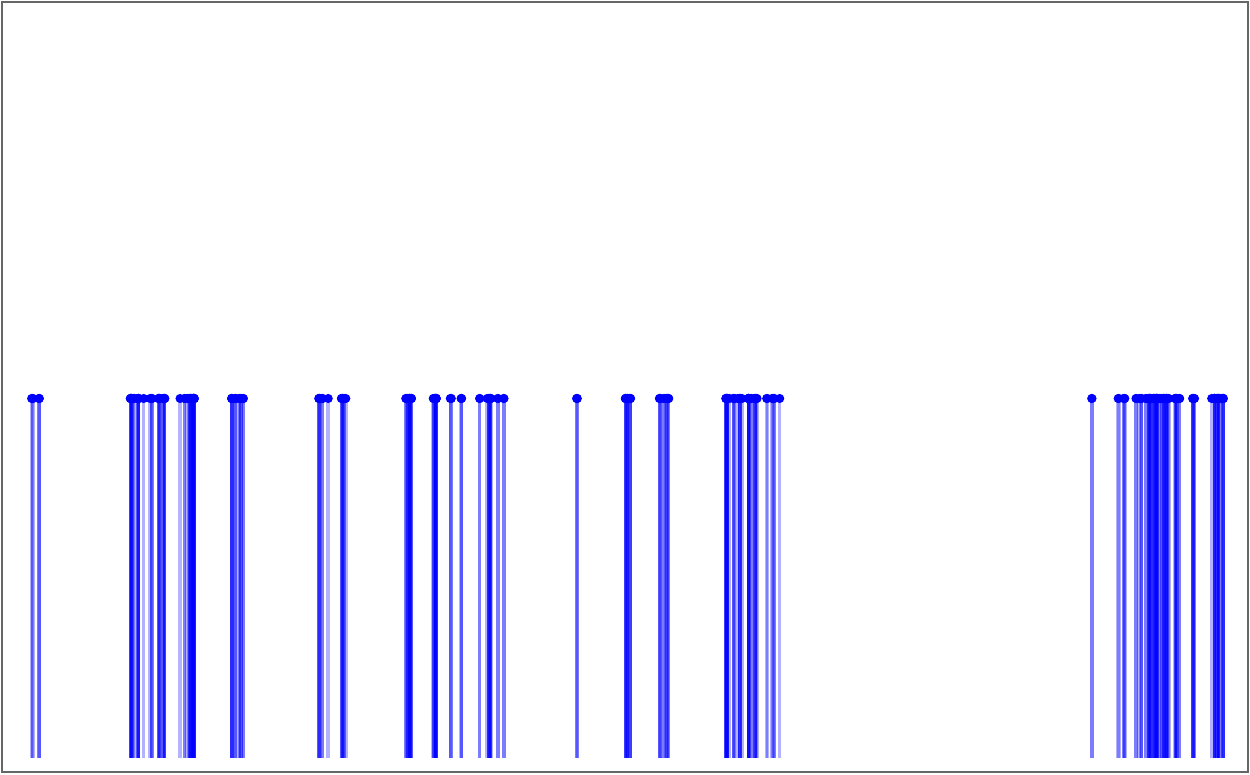}
\includegraphics[width=0.45\linewidth]{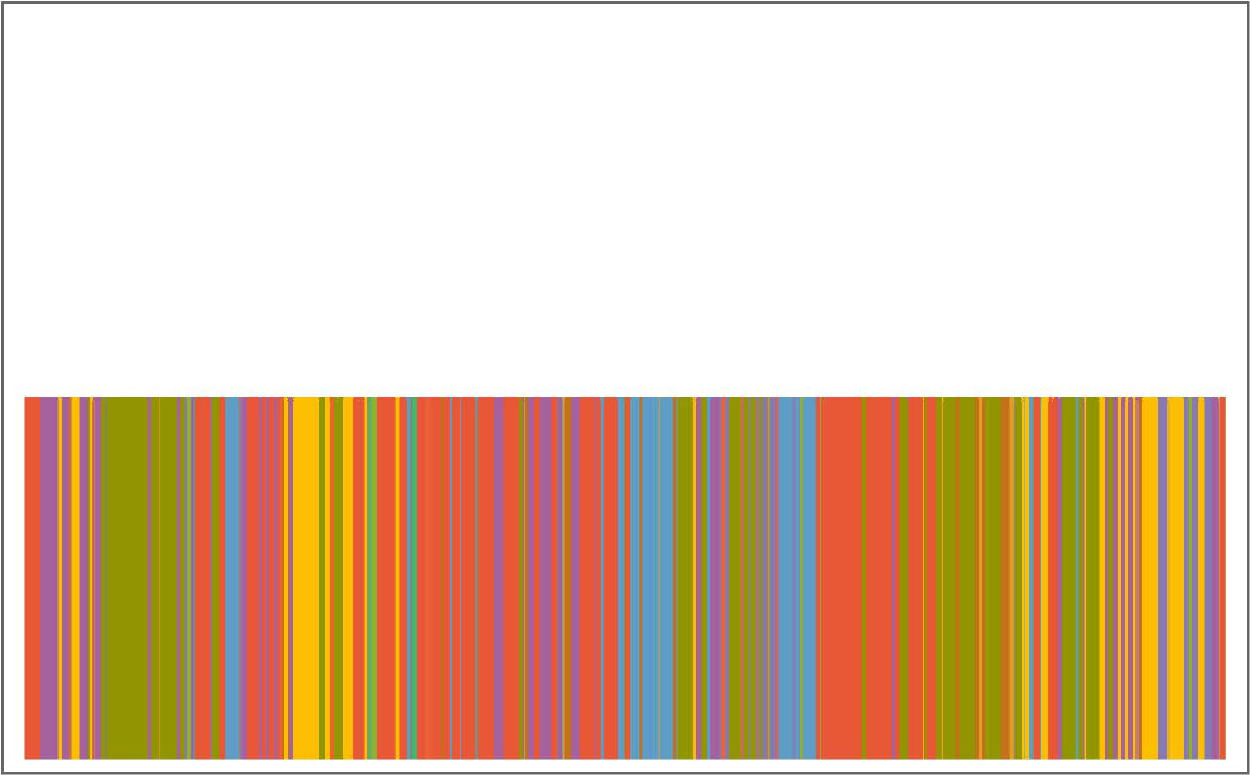}}
\caption{(Color online) (Left panel): Random Cantor set in the local spectrum with $D_{s}=0.63$. (Right panel): Dense global spectrum as a unification of random Cantor sets (shown by different colors) seen in different observation points.}
\label{Fig:Cantor}
\end{figure}	
%%%%%%%%%%%%%%%%%%%%%%%%%%%%%%%%%%%%%%%%%%%%%%%%%%%%%%%%%%%%%%%%%%%%%%%%%%%%%%%%%%%%%%%%%%%%%%%%%%%%%%%%%%%%%%%%%%%%%%%%%%%%%%%%%%%%%%%%%
%%%%%%%%%%%%%%%%%%%%%%%%%%%%%%%%%%%%%%%%%%%%%%%%%%%%%%%%%%%%%%%%%%%%%%%%%%%%%%%%%%%%%%%%%%%%%%%%%%%%%%%%%%%%%%%%%%
\subsection{Random Cantor set in Power-Law banded random matrices}
%%%%%%%%%%%%%%%%%%%%%%%%%%%%%%%%%%%%%%%%%%%%%%%%%%%%%%%%%%%%%%%%%%%%%%%%%%%%%%%%%%%%%%%%%%%%%%%%%%%%%%%%%%%%%%%%%%%
A well-known example of such a situation is a random Cantor set (see Fig.\ref{Fig:Cantor}) which will be considered in detail in the next Section. This type of local spectrum  is realized in the Critical Power-Law Banded Random Matrix (CPLBRM) ensemble \cite{PLBRM} defined as follows. It is an ensemble of random Hermitean matrices $\hat{H}$ with real or complex entries $H_{nm}=\Re H_{nm}+ i \Im H_{nm}$ with statistically independent real and imaginary parts, $\Re H_{nm}=\Re H_{mn}$ and $\Im H_{nm}=-\Im H_{mn}$, which are the Gaussian distributed random variables with zero mean and the variance:
\begin{equation}
\langle |H_{nm}|^{2} \rangle =\frac{1}{1+(n-m)^{2}/b^{2}},
\end{equation}
where the only control parameter is the band-width $0<b<\infty$. It is well-known \cite{Mirlin-rev} that all eigenstates $\psi_{i}(r)$ of such matrices are multifractal, i.e. they are extended but non-ergodic with
\begin{equation}
\sum_{r}|\psi_{i}(r)|^{2q}\sim N^{-D_{q}(q-1)},
\end{equation}
where the $q$-dependent $0<D_{q}<1$ is the eigenfunction multifractal dimension that varies from $0$ to $1$ as the parameter $b$ increases from $b=0$ to $b=\infty$.
%%%%%%%%%%%%%%%%%%%%%%%%%%%%%%%%%%%%%%%%%%%%%%%%%%%%%%%%%%%%%%%%%%%%%%%%%%%%%%%%%%%%%%%%%%%%%%%%%%%%%%%%%%%%%%%%%%%
\subsection{Mini-bands in the local spectrum: Rosenzweig-Porter Random matrix ensemble}
%%%%%%%%%%%%%%%%%%%%%%%%%%%%%%%%%%%%%%%%%%%%%%%%%%%%%%%%%%%%%%%%%%%%%%%%%%%%%%%%%%%%%%%%%%%%%%%%%%%%%%%%%%%%%%%%%%%

%%%%%%%%%%%%%%%%%%%%%%%%%%%%%%%%%%%%%%%%%%%%%%%%%%%%%%%%%%%%%%%%%%%%%%%%%%%%%%%%%%%%%%%%%%%%%%%%%%%%%%%%%%%%%%%%%%%
\begin{figure}[h!]
\center{
\includegraphics[width=0.7\linewidth]{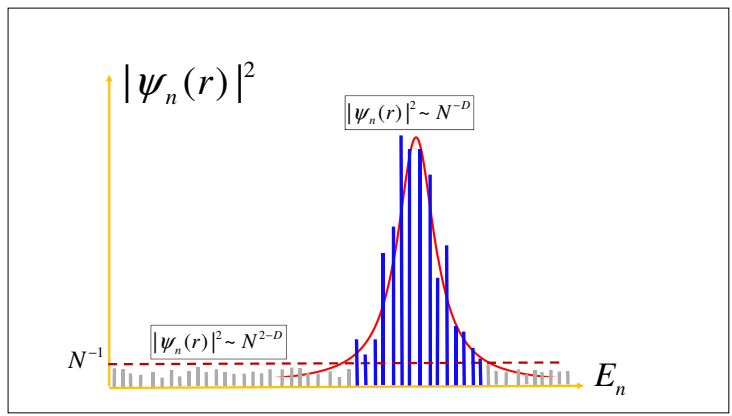}
}
\caption{(Color online) Typical local spectrum of the Rosenzweig-Porter random matrix ensemble in its non-ergodic extended phase. States depicted in blue survive discrimination (set by the red dashed line) and form a mini-band with the Lorentz envelope of $|\psi_{n}(r)|^{2}$ (red solid line). The states shown in gray are discarded. The number of levels in the mini-band $M\sim N^{D}$ is extensive but at $0<D<1$ is much less then the total number of states $N$. The width of the Lorentzian  envelope is $M\delta\sim N^{-(1-D)}$ and tends to zero in the thermodynamic limit.}
\label{Fig:mini-band}
\end{figure}	
%%%%%%%%%%%%%%%%%%%%%%%%%%%%%%%%%%%%%%%%%%%%%%%%%%%%%%%%%%%%%%%%%%%%%%%%%%%%%%%%%%%%%%%%%%%%%%%%%%%%%%%%%%%%%%%%%%%%%%%%%%%%%%%%%%%%%%%%%%%%%
 The difference between the global and the local DoS in the delocalized systems is especially drastic when most of the states are discarded in the discrimination process and the remaining ones form a compact {\it mini-band} (see Fig.\ref{Fig:mini-band}). Such a situation emerges in the Rosenzweig-Porter (RP) random matrix ensemble \cite{RP60,RP15,FacoetiBiroli_RP,Monthus17a, Monthus17b}.
As in the case of CPLBRM the RP random matrices have statistically-independent random Gaussian entries with zero mean. But the variance is bi-modal:
\begin{equation}
\langle |H_{nn}|^{2}\rangle=1,\;\;\;\;\langle |H_{n\neq m}|^{2}\rangle= N^{-\gamma},
\end{equation}
where $0<\gamma<\infty$ is the effective disorder parameter, and $N$ is the size of a matrix.

In the non-ergodic extended phase $1<\gamma<2$ the mini-band in this model contains $M\sim N^{D}$ fractal states of the typical amplitude $|\psi|^{2}\sim N^{-D}$ characterized by the fractal dimension:
\begin{equation}
D_{q}=D=2-\gamma,
\end{equation}
which is independent of $q$ for all $q>1/2$. It is important that inside the mini-band the statistics of local spectrum is Wigner-Dyson \cite{Mehta,Berk,return, SkvorKrav}.
 Thus in the thermodynamic limit $N\rightarrow\infty$ the width of the mini-band $M\delta\sim N^{D-1}$ shrinks to zero but the number of levels in it tends to infinity.
%%%%%%%%%%%%%%%%%%%%%%%%%%%%%%%%%%%%%%%%%%%%%%%%%%%%%%%%%%%%%%%%%%%%%%%%%%%%%%%%%%%%%%%%%%%%%%%%%%%%%%%%%%%%%%%%%%%%%%%%%%%%%%%%%%%%%%%%%%%%%
\section{Regular Cantor set as a spectral counting function}
%%%%%%%%%%%%%%%%%%%%%%%%%%%%%%%%%%%%%%%%%%%%%%%%%%%%%%%%%%%%%%%%%%%%%%%%%%%%%%%%%%%%%%%%%%%%%%%%%%%%%%%%%%%%%%%%%%%%%%%%%%%%%%%%%%%%%%%%%%%%%
In order to better understand the properties of the random Cantor set and its connection with the local spectrum of quantum systems such as CPLBRM,
let us consider the toy model of local spectrum in a form of a regular textbook Cantor set.   Namely, let us consider a set of $M=N^{D_{s}}$ delta-functions $\delta(E-E_{n})$ with the equal weight $|\psi_{n}(r)|^{2}=N^{-D_{s}}$. Such a model obeys the completeness condition:
\begin{equation}
\sum_{n=1}^{M}|\psi_{n}(r)|^{2}=1.
\end{equation}
The points $E_{n}$ in the segment $[0,1]$ which represent the eigenvalues in this toy model are organized in generations as follows (see Eq.(\ref{gener})). The first generation consists of two points $1/3$ and $2/3$ which are the ends of the central $1/3$-segment. The  second generation  are the ends of the central $1/3$-segment of the remaining two segments. This procedure is repeated until the minimal spacing $3^{-n}$ becomes $N^{-1}$.
\newpage
\begin{eqnarray}\label{gener}
  \frac{1}{3}\;&&\;\frac{2}{3} \\ \nonumber
 \frac{1}{9}\;\;\frac{2}{9}\;\;&&\;\;\;\frac{7}{9}\;\;\frac{8}{9} \\ \nonumber
 \frac{1}{27}\;\frac{2}{27}\;\;\frac{7}{27}\,\frac{8}{27}&&\frac{19}{27}\;\frac{20}{27}\;\;\frac{25}{27}\;\frac{26}{27}  \\ \nonumber
................&&................\\ \nonumber
\end{eqnarray}
The entire spectrum is the unification of all $n$ generations.
  The total number of levels in the spectrum is the sum over generations:
\begin{equation}\label{M}
M=\sum_{i=1}^{n} 2^{i}=2 (2^{n}-1)\approx 2^{n+1}=2N^{\frac{\ln2}{\ln3}}.
\end{equation}
According to Eq.(\ref{D_s}) this implies:
\begin{equation}\label{ln2_ln_3}
D_{s}=\frac{\ln2}{\ln3}
\end{equation}
Next, we notice that the number of times the spacing $s_{m}= 3^{-m}$ occurs is $2^{m-1}$.   Thus the probability $p(s_{m})$ to have spacing $s_{m}$ appears to be a power-law:
\begin{equation}
p(s_{m})=\frac{2^{m-1}}{2^{n}}=\frac{1}{2}\left(\frac{\delta}{s_{m}} \right)^{D_{s}}.
\end{equation}
 This corresponds to the power-law probability {\it density}, or the {\it spacing distribution function} $P(s)$, for the regular Cantor set:
\begin{equation}\label{P_s}
P(s)=\left\{ \matrix {A\,\delta^{D_{s}}\,s^{-(D_{s}+1)}, & (1>s>\delta=N^{-1})\cr
0, & \mbox{otherwise}}\right.,
\end{equation}
where $A$ is the normalization constant and $D_{s}$ is given by Eq.(\ref{ln2_ln_3}). For $s<\delta$ and for $s>1$  $P(s)$ is zero. One can easily see that Eq.(\ref{delta_typ-av}) is valid for this distribution.

This form of $P(s)$ will be the key starting point to define the {\it random Cantor set} in the next Section.
%%%%%%%%%%%%%%%%%%%%%%%%%%%%%%%%%%%%%%%%%%%%%%%%%%%%%%%%%%%%%%%%%%%%%%%%%%%%%%%%%%%%%%%%%%%%%%%%%%%%%%%%%%%%%%%%%%%
  \subsection{Construction of a regular Cantor counting function}
\begin{figure}[t]
\center{
\includegraphics[width=0.4\linewidth]{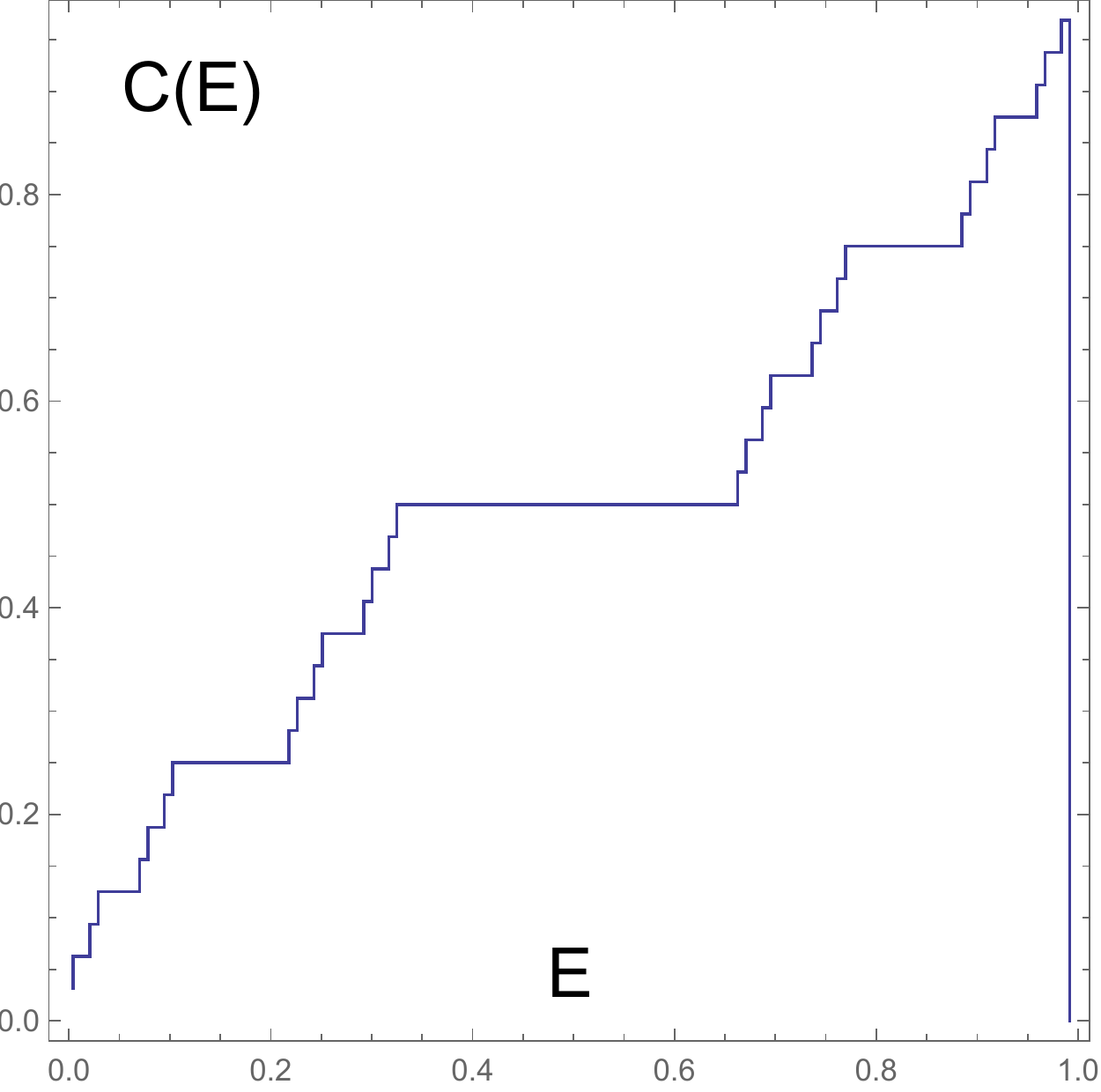}
}
\caption{(Color online) Cantor counting function}
\label{Fig:reg_Cantor}
\end{figure}

%%%%%%%%%%%%%%%%%%%%%%%%%%%%%%%%%%%%%%%%%%%%%%%%%%%%%%%%%%%%%%%%%%%%%%%%%%%%%%%%%%%%%%%%%%%%%%%%%%%%%%%%%%%%%%%%%%%
Now let us  construct the {\it counting function} $C(E)$ corresponding to the regular Cantor set defined by Eq.(\ref{gener}):
\begin{equation}\label{count_Cantor}
C(E)=M^{-1}\sum_{i=1}^{M}\theta(E-E_{i}),
\end{equation}	
where $\theta(x)$ is the Heaviside step function, $E_{i}$ is the ordered set of 'eigenvalues' $E_{n}$ defined by Eq.(\ref{gener}) and $M\sim N^{D_{s}}$ is the total number of levels, Eq.(\ref{M}). The so defined {\it Cantor counting function} is plotted in Fig.\ref{Fig:reg_Cantor}.
%%%%%%%%%%%%%%%%%%%%%%%%%%%%%%%%%%%%%%%%%%%%%%%%%%%%%%%%%%%%%%%%%%%%%%%%%%%%%%%%%%%%%%%%%%%%%%%%%%%%%%%%%%%%%%%%%%%
%%%%%%%%%%%%%%%%%%%%%%%%%%%%%%%%%%%%%%%%%%%%%%%%%%%%%%%%%%%%%%%%%%%%%%%%%%%%%%%%%%%%%%%%%%%%%%%%%%%%%%%%%%%%%%%%%%%
%%%%%%%%%%%%%%%%%%%%%%%%%%%%%%%%%%%%%%%%%%%%%%%%%%%%%%%%%%%%%%%%%%%%%%%%%%%%%%%%%%%%%%%%%%%%%%%%%%%%%%%%%%%%%%%%%%%
\subsection{Convolution of two Cantor counting functions and the LDoS correlation function}
From the definition of Cantor counting function it follows that the derivative $\partial_{E}C(E)$ is the set of $\delta$-functions that mimics the LDoS. For random systems one is interested in the {\it correlation function}
of two LDoS defined by Eq.(\ref{local_DoS}):
\begin{equation}\label{Kom}
K(\omega)=\langle\rho(E+\omega/2;r)\,\rho(E-\omega/2;r) \rangle.
\end{equation}	
In terms of the counting functions:
\begin{equation}\label{C-gen}
C(E;r)=\sum_{n}|\psi_{n}(r)|^{2}\,\theta(E-E_{n}),
\end{equation}
we have:
\begin{equation}\label{K_om_random}
K(\omega)=-4\frac{d^{2}}{d\omega^{2}}\,\langle C(E+\omega/2;r)\,C(E-\omega/2) \rangle
\end{equation}
In our toy model with the regular Cantor counting function $C(E)$ instead of the random $C(E;r)$ we replace the ensemble averaging by the averaging over the spectrum:
\begin{equation}\label{K-Cantor}
K_{{\rm Cantor}}(\omega)=-4\frac{d^{2}}{d\omega^{2}}\,\int_{\frac{\omega}{2}}^{1-\frac{\omega}{2}} C(E+\omega/2)C(E-\omega/2)\,dE,
\end{equation}
where $C(E)$ is given by Eq.(\ref{count_Cantor}).
%%%%%%%%%%%%%%%%%%%%%%%%%%%%%%%%%%%%%%%%%%%%%%%%%%%%%%%%%%%%%%%%%%%%%%%%%%%%%%%%%%%%%%%%%%%%%%%%%%%%%%%%%%%%%%%%%%%%%%%%%%%%%%%%%%%%%%%%%%%%%
\begin{figure}[t]
\center{
\includegraphics[width=0.4\linewidth]{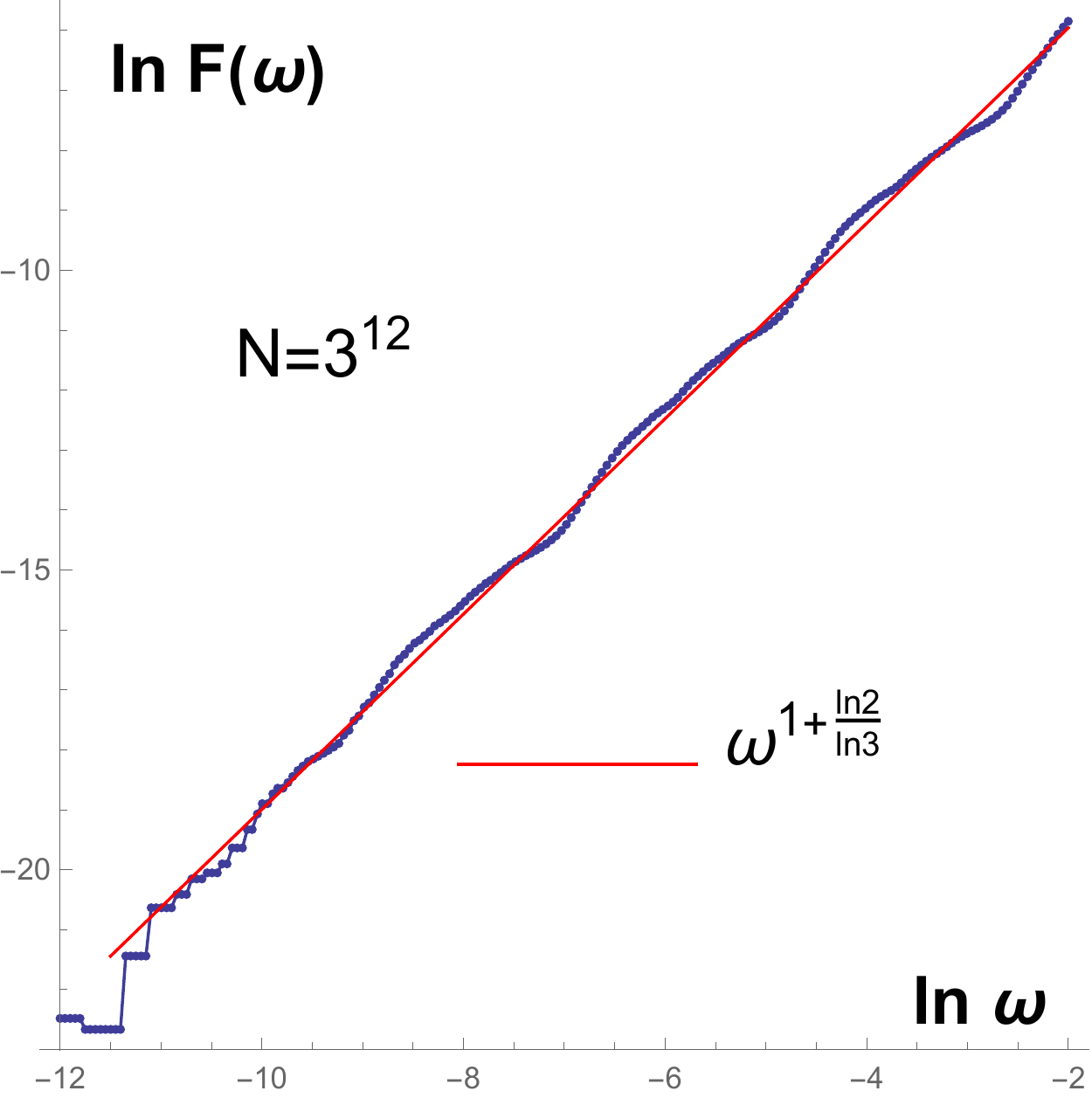}
}
\caption{(Color online) $C(E+\omega/2)C(E-\omega/2)-C(E)^{2}$ integrated over $E$ as in Eq.(\ref{K-Cantor}) for the Cantor set with $n=12$ generations.}
\label{Fig:C-integr}
\end{figure}
%%%%%%%%%%%%%%%%%%%%%%%%%%%%%%%%%%%%%%%%%%%%%%%%%%%%%%%%%%%%%%%%%%%%%%%%%%%%%%%%%%%%%%%%%%%%%%%%%%%%%%%%%%%%%%%%%%%%%%%%%%%%%%%%%%%%%%%%%%%%%
In Fig.\ref{Fig:C-integr} the averaging over the spectrum for $F(\omega)=C(E+\omega/2)C(E-\omega/2)-C(E)^{2}$ is done numerically for the Cantor set with $n=12$ generations. One can see that the result is well described by a power law $F(\omega)\propto \omega^{1+D_{s}}$. We conjecture that in the limit $N=3^{n}\rightarrow\infty$ this power law becomes exact, so that {\it down to $\omega=0$} we have:
\begin{equation}\label{K_Cantor}
K_{{\rm Cantor}}(\omega)\propto -4\frac{d^{2}F(\omega)}{d\omega^{2}}\sim \left(\frac{1}{\omega}\right)^{1-D_{s}},\;\;\;\; \left(D_{s}=\frac{\ln2}{\ln3}\right).
\end{equation}
The regular Cantor set is the prototype of the  {\it singular continuous spectrum} defined rigorously in mathematical literature \cite{Dunford}. We see, therefore, that the 'smoking gun' of  the local analogue of such a spectrum, the $l$-singular continuous spectrum, in the limit $N\rightarrow\infty$, is the singular behavior of the LDoS correlation function, Eq.(\ref{Kom}), at $\omega\rightarrow 0$.
%%%%%%%%%%%%%%%%%%%%%%%%%%%%%%%%%%%%%%%%%%%%%%%%%%%%%%%%%%%%%%%%%%%%%%%%%%%%%%%%%%%%%%%%%%%%%%%%%%%%%%%%%%%%%%%%%%%%%%%%%%%%%%%%%%%%%%%%%%%%%%%%%%%%%%%%%
\section{Random Cantor set}
%%%%%%%%%%%%%%%%%%%%%%%%%%%%%%%%%%%%%%%%%%%%%%%%%%%%%%%%%%%%%%%%%%%%%%%%%%%%%%%%%%%%%%%%%%%%%%%%%%%%%%%%%%%
\subsection{Random Cantor set and its counting function}
%%%%%%%%%%%%%%%%%%%%%%%%%%%%%%%%%%%%%%%%%%%%%%%%%%%%%%%%%%%%%%%%%%%%%%%%%%%%%%%%%%%%%%%%%%%%%%%%%%%%%%%%%%%
%%%%%%%%%%%%%%%%%%%%%%%%%%%%%%%%%%%%%%%%%%%%%%%%%%%%%%%%%%%%%%%%%%%%%%%%%%%%%%%%%%%%%%%%%%%%%%%%%%%%%%%%%%%%%%%%%%%
\begin{figure}[tbh]
\center{
\includegraphics[width=0.30\linewidth]{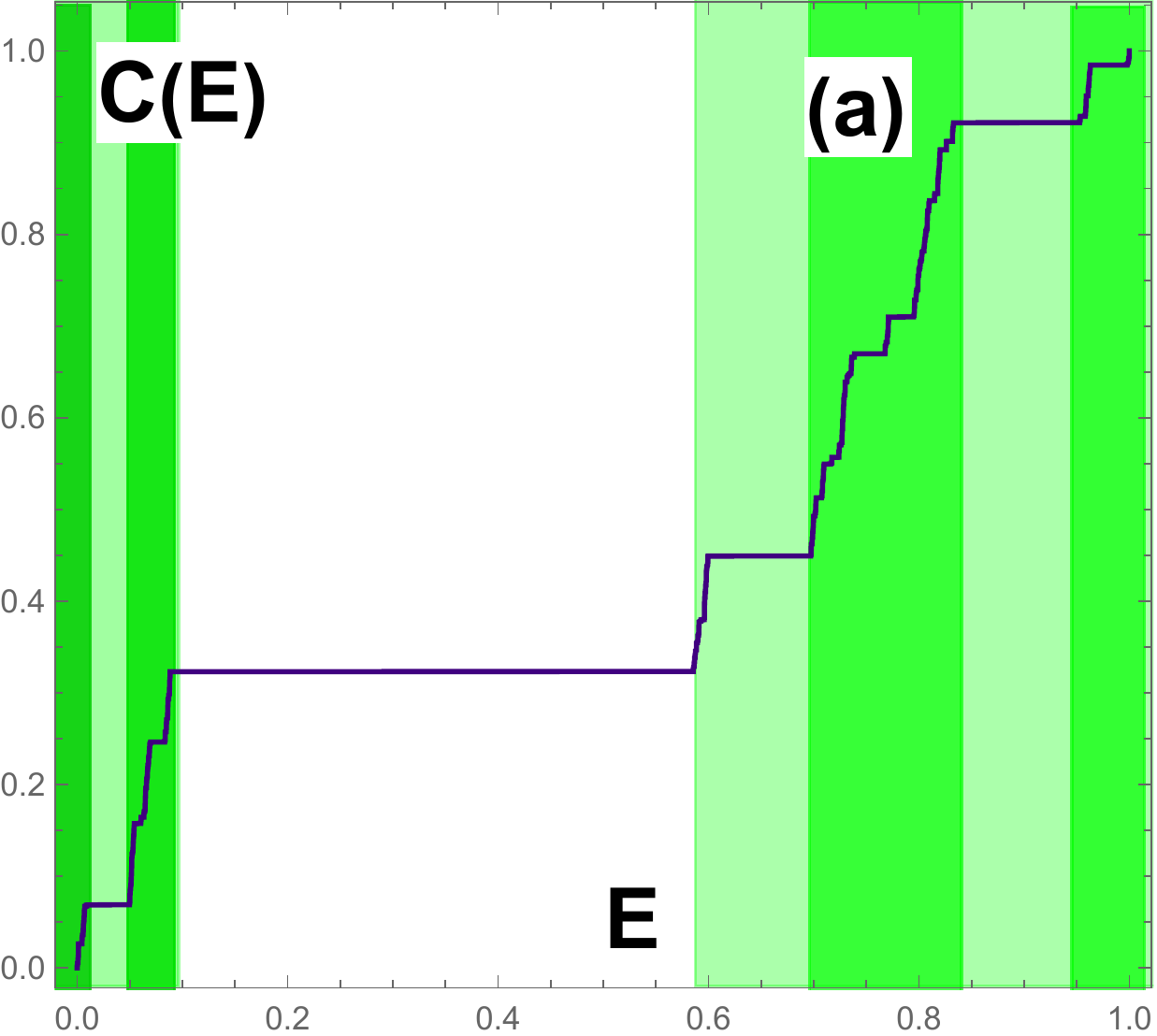}
\includegraphics[width=0.30 \linewidth]{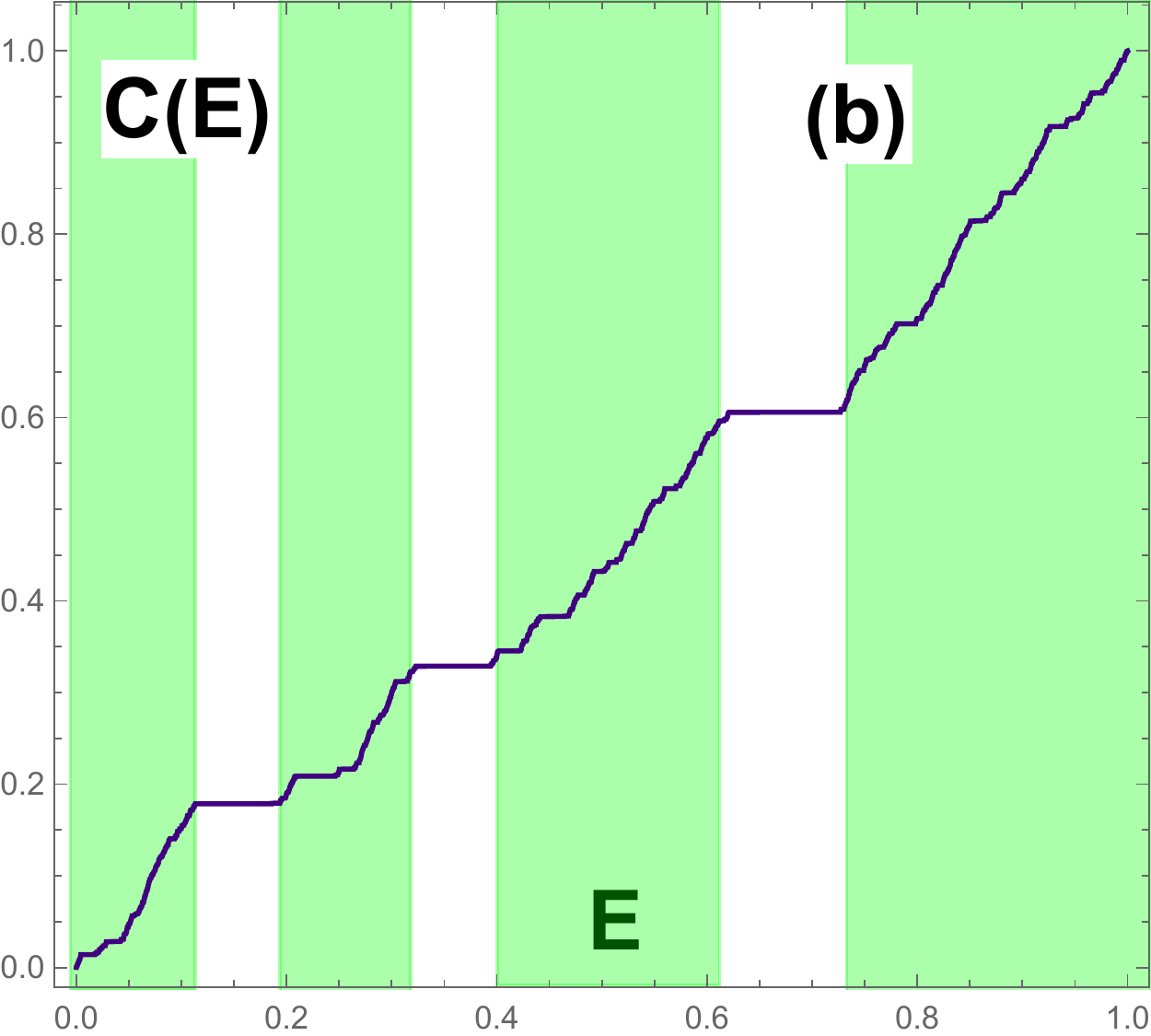}
\includegraphics[width=0.30\linewidth]{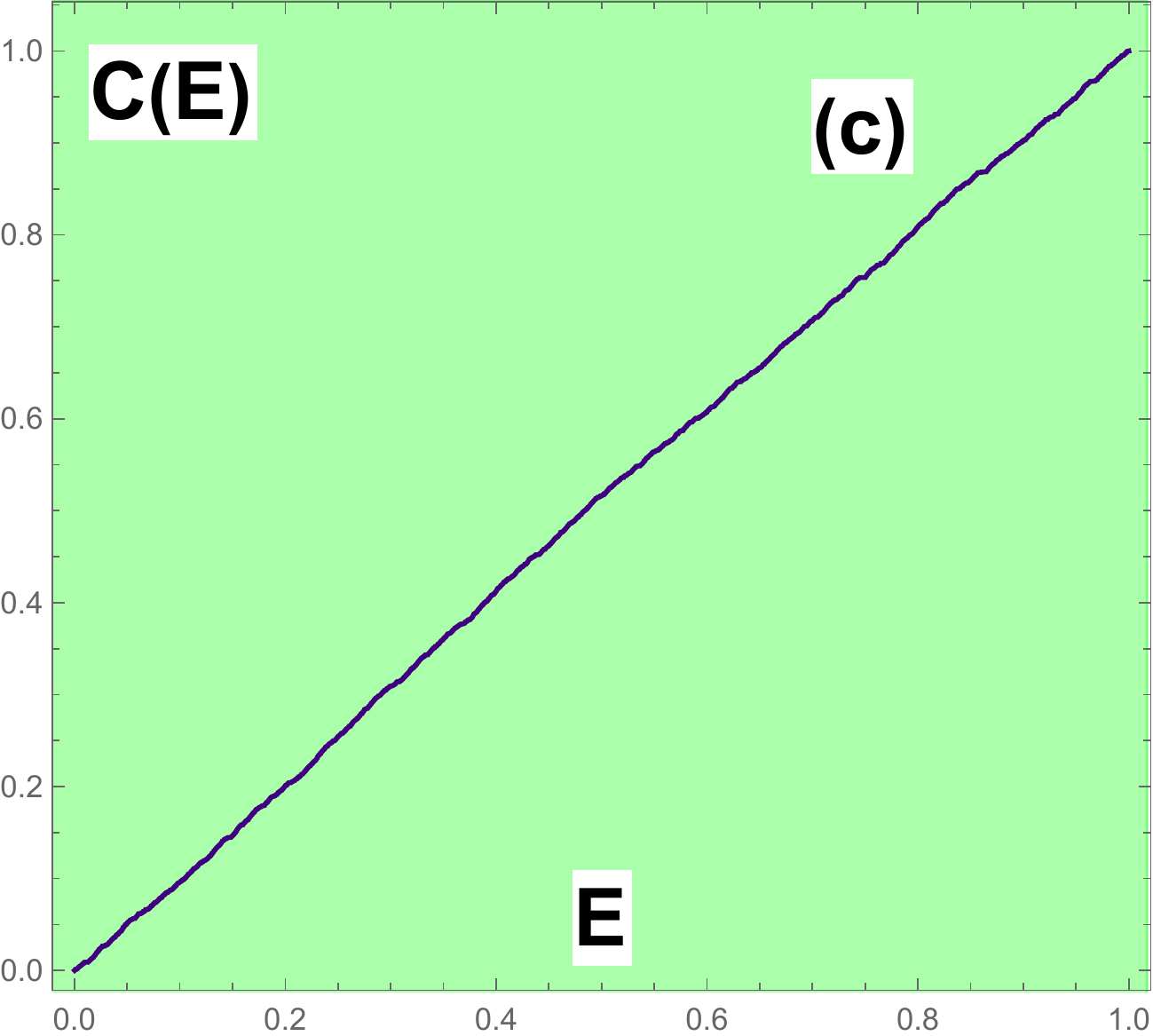}
\includegraphics[width=0.30\linewidth]{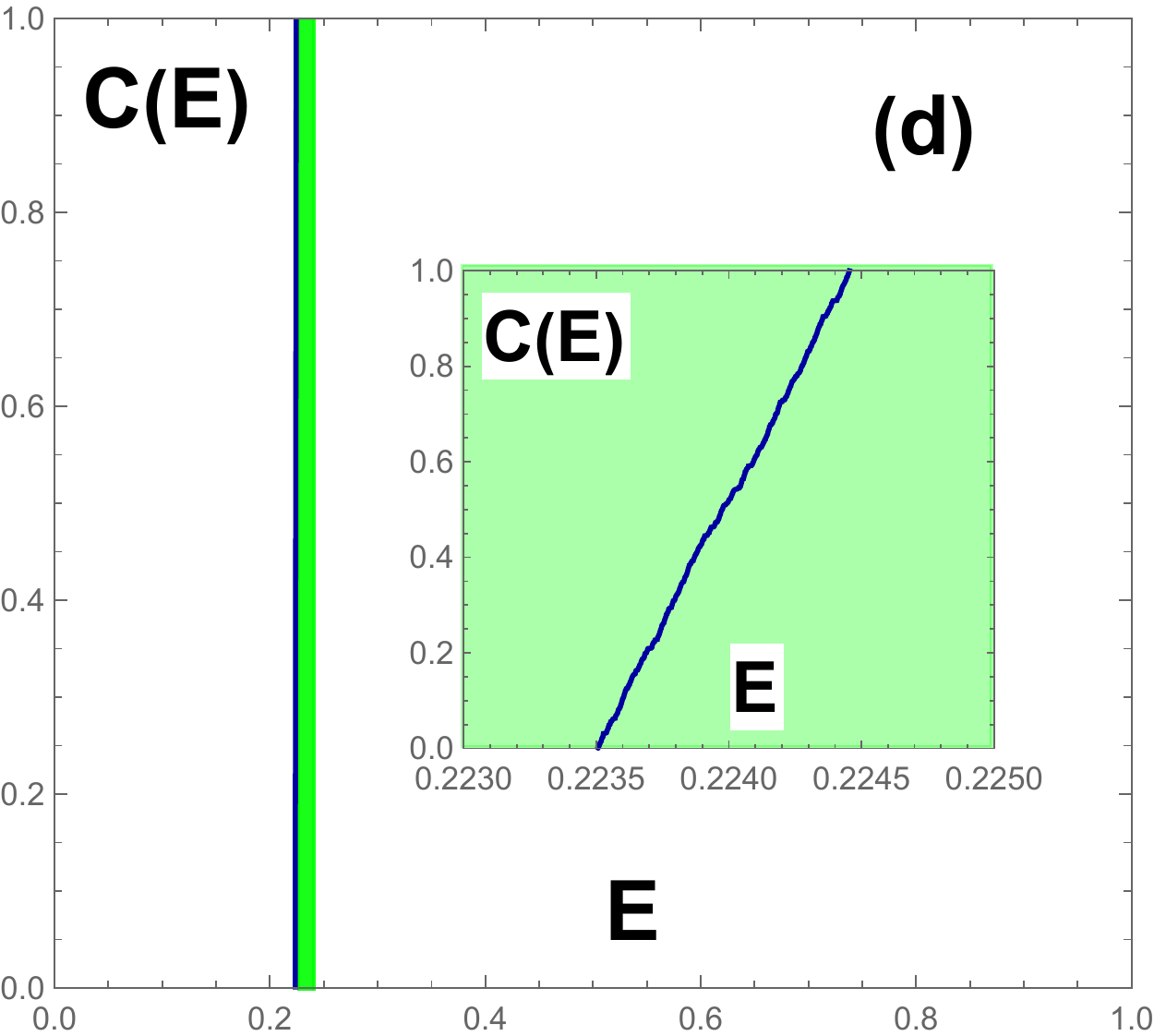}
}
\caption{(Color online)The counting function $C(E)$ for  a set of $M=N^{D}$ levels with the power-law spacing distribution, Eq.(\ref{P_s}),  characterized by: (a) $D=D_{s}=0.63$;   (b) $D=D_{s}=1$;
(c) $D=1, D_{s}=2$; (d) $D=0.5, D_{s}=2$. In the random Cantor set (a) the gaps (plateaus) between the mini-bands (green strips) at any scale $\Delta$ in the hierarchy (shown by increasing intensity of green strips) are of the order of the typical width of   mini-bands. In the case (b) the  gaps between the strings of levels  shown by green strips, are smaller than the typical width of the strings (a {\it sub-dense} spectrum).  In this case the local spectrum lacks a hierarchy of mini-bands.  In the case (c) the spectrum is dense with no large gaps at all. In the case (d)  an isolated narrow mini-band appears with the zoomed counting function of the type (c) shown in the inset. A similar isolated mini-band appears in all the cases when $D<D_{s}$ with the spectrum inside the mini-band  being a Cantor  set (a)
if $D_{s}<1$ or it is sub-dense as in (b) if $D_{s}=1$.
%In the case (e) where $D>D_{s}$ the local spectrum consists of several mini-bands with the Cantor set  spectrum (shown in the inset) inside each of them.
}
\label{Fig:C_random_Cantor}
\end{figure}
The power law, Eq.(\ref{P_s}) for the  spacing distribution in the regular Cantor set can be used to define a {\it random Cantor set} which is a natural generalization of a regular Cantor set to describe the singular continuous spectrum of random Hamiltonians.

Let us consider a model of local spectrum in which all $s_{n}=E_{n+1}-E_{n}$ are:
\begin{itemize}
\item
independent random quantities,
\item
distributed identically according to the power-law Eq.(\ref{P_s}) with some $1>D_{s}>0$.
\end{itemize}
An example of the counting function $C(E)$ drawn from such a model of random spectra with $D_{s}=0.63$ is shown in Fig.\ref{Fig:C_random_Cantor}(a). Like in the regular Cantor set, this counting functions consists of the hierarchy of plateaus and staircases between them (denoted by green stripes of increasing intensity as the scale decreases). The length  of the staircases (the width of the green stripes) at all scales is of the order of the length of the plateaus at the corresponding scales.

%%%%%%%  %%%%%%%%%%%%%%%%%%%%%%%%%%%%%%%%%%%%%%%%%%%%%%%%%%%%%%%%%%%%%%%%%%%%%%%%%%%%%%%%%%%%%%%%%%%%%%%%%%%%
\subsection{Correlation function of two random Cantor counting functions}\label{subsect:corr-funct}
%%%%%%%%%%%%%%%%%%%%%%%%%%%%%%%%%%%%%%%%%%%%%%%%%%%%%%%%%%%%%%%%%%%%%%%%%%%%%%%%%%%%%%%%%%%%%%%%%%%%%%%%%%%
Now we compute the ensemble average of the product of a random Cantor counting function and the same counting function shifted by $\omega$ in the energy space. Such an average corresponds to the  LDoS correlation function,
provided that the local spectrum (after discrimination) is a random Cantor set. We demonstrate that the result of such averaging for a model with $D_{s}<1$ reproduces the LDoS correlation function in many critical systems (including the Anderson models at the localization transition and CPLBRM) just corroborating  the conjecture that the local spectrum of the corresponding quantum critical systems is a random Cantor set.

Because of  independently fluctuating  and identically distributed spacings $s_{j}$ in our model of a random Cantor set one can easily compute the correlation function   $K(\omega)$ given by Eq.(\ref{K_om_random}). First we note that since the averaged LDoS and its correlation function $K(\omega)$ for the random Cantor set are independent of $E$ everywhere in the interval $[0,1]$ one may integrate Eq.(\ref{K_om_random})[where $C(E;r)$ is given by Eq.(\ref{count_Cantor}) with random $E_{i}$] over $E$ in this interval and represent $K(\omega)$ as follows:
\begin{equation}\label{KK}
K(\omega)=M^{-2}\sum_{i,i'=1}^{M}\delta(\omega-E_{i}+E_{i'}),
\end{equation}
where $M\leq N$  is the total number of levels $E_{i}$ in our model of {\it local} spectrum, and
\begin{equation}
E_{i}-E_{i'}=\sum_{j=i'}^{i}s_{j}.
\end{equation}
The function $K(\omega)$ is normalized:
\begin{equation}\label{normm}
\int K(\omega)\,d\omega =1.
\end{equation}
Furthermore, for $M\gg 1$ its {\it regular} part corresponding to $i=i'$ in Eq.(\ref{KK}) also obeys Eq.(\ref{normm}) with the accuracy $O(1/M)$.

Next, using $\langle e^{-i\tau \sum_{j=i'}^{i} s_{j}}\rangle =\langle e^{-i\tau\, s}\rangle^{i-i'}$ that follows from the statistical independence of $s_{j}$, and summing the geometric series one obtains   for the Fourier transform of $K(\omega)$:
\begin{eqnarray}\label{K-q}
\tilde{K}(\tau)&=&M^{-2}\left[M+2 \sum_{i'=1}^{M}\sum_{i=i'+1}^{M}\,   \Re \langle e^{-i\tau \sum_{j=i'}^{i} s_{j}}\rangle\right]\\ \nonumber &=& M^{-1}+2 M^{-2}\Re\left[ \left(M-1-\frac{g-g^{M}}{1-g}\right)\,\frac{g}{1-g}\right].
\end{eqnarray}
 In Eq.(\ref{K-q}) we denote $g\equiv g(t=i\tau)$ where on a real axis $t>0$:
\begin{equation}\label{q}
g(t)=\langle e^{-t\,s}\rangle = \int_{0}^{\infty} e^{-t\,s}\,P(s)\,ds.
\end{equation}
By normalization of the probability distribution $P(s)$
  one has on the real axis of $t$:
\begin{equation}\label{properties}
g(0)=1,\;\;\;\;\;\; 0<g(t)\leq 1.
\end{equation}
The function $\tilde{K}(\tau)$ gives the mean return/survival probability:
\begin{equation}
\tilde{K}(\tau)=\langle |\Psi(\tau;r)|^{2}\rangle,
\end{equation}
where initially at $\tau=0$ the wave function $\Psi(t=0,r)$ was non-zero only at a point $r$.

The spectral form-factor is given by an inverse Fourier transform of $\tilde{K}(\tau)$:
\begin{equation}\label{inverse}
K(\omega)=\int_{-\infty}^{+\infty}\tilde{K}(\tau)\,e^{i\omega \tau}\,\frac{d\tau}{2\pi}.
\end{equation}
The function $g(t)$ that corresponds to the distribution Eq.(\ref{P_s}) at $0<D_{s}<1$ has  a brunch-cut  singularity at $t=0$:
\begin{equation}\label{q-expan}
g(t)= 1- c\, (\bar{t})^{D_{s}},
\end{equation}
where $\bar{t}=t\delta=t/N<<1$ and $c$ is a positive constant of order 1.
At $D_{s}=1$ one obtains:
\begin{equation}\label{q_D_s_1}
g(t)=1+\bar{t}\,\ln\bar{t},
\end{equation}
while for $2>D_{s}>1$ a  linear in $\bar{t}$ term appears:
\begin{equation}\label{q_D_s_great_1}
g(t)=1-c_{1}\bar{t}-c_{2}\bar{t}^{D_{s}}.
\end{equation}
In general a regular $\bar{t}^{m}$ expansion with an integer $m$ holds for the first $m=[D_{s}]$ terms, where $[...]$ denotes an integer part.

At large enough $M$ the term $\propto g^{M}$ in Eq.(\ref{K-q}) (i.e. the boundary term in summation of the geometric series) can be omitted.   In this case one obtains:
\begin{equation}\label{K_small_om}
\tilde{K}(\tau)\approx 2M^{-1}\Re\,(1-g(i\tau))^{-1},
\end{equation}
which for a random Cantor set ($D_{s}<1$ and $M=N^{D_{s}}$) results in:
\begin{equation}\label{K-omega}
K(\omega)\sim  \frac{M^{-1}\delta^{-D_{s}}}{{\bf \omega}^{{\bf 1-D_{s}}}}\sim \frac{1}{{\bf \omega}^{{\bf 1-D_{s}}}},
\end{equation}
i.e. the same power law, Eq.(\ref{K_Cantor}), as for the regular Cantor set but with an arbitrary $0<D_{s}<1$.

It is remarkable that   Eq.(\ref{K-omega}) at $D_{s}=D_{2}=d_{2}/d$ (where $d_{2}$ is the multi-fractal dimension of the critical wave function and $d$ is the dimensionality of the lattice)   reproduces exactly the Chalker's ansatz first established by Chalker and Daniel for the Integer Quantum Hall effect  at the center of Landau band  \cite{Chalker_Daniel} and later on found true for the $d$-dimensional Anderson model at the mobility edge $E=E_{c}$ \cite{Chalker}\footnote{The critical region near the mobility edge or the center of Landau band $E=E_{c}$ is defined so as the critical length $\xi\propto |E-E_{c}|^{-\nu}$ is larger than the system size $L=N^{\frac{1}{d}}$. Thus in the critical energy window $|E-E_{c}|<N^{-\frac{1}{\nu d}}$ there are $N^{1-\frac{1}{\nu d}}$ critical states, which by Harris criterion \cite{Harris} $\nu d >2$ tends to infinity in the thermodynamic limit. } and for the Power-Law Banded Random Matrix ensemble \cite{PLBRM,KrMuttalib97,CueKrav07,OssKr10}. Thus we can argue that in all these cases the local spectrum (after discrimination described in Sec.\ref{discr} )  is a random Cantor set.

 However, at $D_{s}=1$  one obtains from Eqs.(\ref{K_small_om}),(\ref{q_D_s_1})   a slowly decreasing
\begin{equation}\label{K_D_s_1}
K(\omega)\sim M^{-1}\delta^{-1} \,\left(\ln\frac{\omega}{\delta}
\right)^{-1},
\end{equation}
 rather than a constant. Indeed, we see from Fig.\ref{Fig:C_random_Cantor}(b) that the corresponding counting function does contain the plateaus and staircases, albeit the length of the staircases is now larger than that of the plateaus. The corresponding local spectrum is no longer a random Cantor set, neither it is dense as for ergodic systems. The model, Eq.(\ref{P_s}), with $D_{s}=1$ represents a border-line case in between of a random Cantor set and a dense spectrum (which we refer to as a {\it sub-dense} spectrum). Its Hamiltonian realization is not known so far to the best of our knowledge.

It is only at $D_{s}>1$ where the constant $K(\omega)$ at $\omega\gg\delta$ emerges. The point is that with taking the boundary term $g^{M}$ into account in Eq.(\ref{K-q}) one should modify Eq.(\ref{K_small_om}) as follows:
\begin{equation}\label{K_small_om_bis}
\tilde{K}(\tau)\approx 2M^{-1}\Re\,(1-g(i\tau)+\eta)^{-1},
\end{equation}
where $\eta= M^{-1}\rightarrow +0$.

Then at $D_{s}>1$ one obtains from Eq.(\ref{q_D_s_great_1}):
\begin{equation}\label{K_tau_D_gr_1}
\tilde{K}(\tau)\sim \Re\,\frac{M^{-1}}{i\tau\delta+\eta+c_{2}\,(i \tau \delta)^{D_{s}} }.
\end{equation}
At $D_{s}>1$, in the leading approximation in $\tau\delta\ll 1$  Eq.(\ref{K_tau_D_gr_1}) results in:
\begin{equation}\label{K_D_s_gr_1}
\tilde{K}(\tau)\sim \Re\,\frac{M^{-1}}{i\tau\delta+\eta}\rightarrow \pi\,M^{-1}\delta^{-1}\, \delta(\tau),
\end{equation}
which implies a constant
\begin{equation}\label{K_om_erg}
K(\omega)\sim M^{-1}\delta^{-1},
\end{equation}
 at $\omega\gg \delta$, like for a dense local spectrum in the Wigner-Dyson RMT. The corresponding typical counting function is shown in Fig.\ref{Fig:C_random_Cantor}(c) and it does not exhibit plateaus which correspond to anomalously large gaps in the local spectrum. This example shows that the case $D_{s}>1$ in Eq.(\ref{P_s}) is essentially the same as for a fast decreasing $P(s)$ [like the Poisson or the Wigner-Dyson $P(s)$ in the global spectrum] without fat tails. Thus this local spectrum is {\it dense}, the same as the global spectrum.
%%%%%%%%%%%%%%%%%%%%%%%%%%%%%%%%%%%%%%%%%%%%%%%%%%%%%%%%%%%%%%%%%%%%%%%%%%%%%%%%%%%%%%%%%%%%%%%%%%%%%%%%%%%%%%%%%%%%%%%%%%%%%%%%%%%%%%%%%%%%%
\section{An isolated mini-band}
%%%%%%%%%%%%%%%%%%%%%%%%%%%%%%%%%%%%%%%%%%%%%%%%%%%%%%%%%%%%%%%%%%%%%%%%%%%%%%%%%%%%%%%%%%%%%%%%%%%%%%%%%%%%%%%%%%%%%%%%%%%%%%%%%%%%%%%%%%%%%
%It is instructive to see what happens to $K(\omega)$ in the limit $D_{s}\rightarrow 1$. If the number of levels %remains to be $M=N^{D_{s}}$ the ergodic limit, Eq.(\ref{K_om_erg}),  will be reached at $D_{s}=1+\epsilon $, %$(\epsilon\rightarrow +0)$.
For a random Cantor set we considered the model in which the number of local levels $M\sim N^{D_{s}}$, where $D_{s}$ controls the power-law distribution Eq.(\ref{P_s}).
However, this is not the only possibility but a very particular case.  Indeed, comparing Eqs.(\ref{count_Cantor}) and (\ref{C-gen}) we see that the number of levels $M$ is in fact related with the multifractal dimension of an eigenfunction. Indeed, in the simplified case (realized in the Rosenzweig-Porter RMT) when all multifractal dimensions $D_{q}=D$ are identical for $q\geq 1/2$, one may replace, by the order of magnitude, $|\psi_{n}(r)|^{2}\rightarrow N^{-D}$, so that $M$ in Eq.(\ref{count_Cantor}) must be $M\sim N^{D}$ in order to reproduce Eq.(\ref{C-gen}), while $D_{s}$ is yet another independent parameter of the model. In a generic multifractal case we have in $K(\omega)$:
\begin{equation}\label{M-D}
M\sim N^{D_{2}},
\end{equation}
as the correlation function $K(\omega)$ is second order in $|\psi|^{2}$.

The question is: What will happen if $D_{2}<D_{s}$? The answer is known for the Rosenzweig-Porter RMT in its fractal phase where $K(\omega)$ has a simple Lorenzian shape \cite{RP15,return,Monthus17a,Monthus17b}.
  It is nearly independent of $\omega$ for $\delta\ll \omega\ll\Gamma$, where $\Gamma\sim N^{-1+D}\gg\delta$ is the width of the mini-band (see Fig.\ref{Fig:mini-band}), and  falls down as $\Gamma/\omega^{2}$ at $\omega\gg\Gamma$. Can we reproduce this behavior and find the correct scaling of $\Gamma\sim N^{-(1-D)}$ with $N$ from the above model of independently fluctuating spacings with $D_{2}<D_{s}$? As a matter of fact, the answer to this question is affirmative.

Indeed, consider the case $D_{2}<D_{s}$ in Eqs.(\ref{M-D}),(\ref{P_s}) of our model of independent  identically distributed spacings. Then there is a region  of $\omega$ where $M\ln(g(t\sim 1/\omega))\lesssim 1$:
\begin{eqnarray}\label{region}
  && N^{-1+\frac{D_{2}}{D_{s}}}\lesssim \omega \ll 1, \;\;\; \;\;D_{s}<1,\\ \nonumber
	&&  N^{-1+D_{2}}\ln(N^{D_{2}})\lesssim\omega \ll 1, \;\;\; \;\;D_{s}= 1,\\ \nonumber
	&& N^{-1+D_{2}}\lesssim \omega \ll 1, \;\;\; \;\;D_{s}\geq 1.
\end{eqnarray}
In this region the boundary term $g^{M}$ in the summation of geometric series, Eq.(\ref{K-q}), cannot be neglected.

 A thorough inspection of  Eq.(\ref{K-q}) in the region Eq.(\ref{region})  gives the following result:
\begin{equation}\label{K_tail}
  \tilde{K}(\tau)\approx 2\left(1-\frac{M}{3}\,\Re (1-g(i\tau)) \right).
\end{equation}

The $\tau$-independent term in Eq.(\ref{K_tail}) results in a delta-function in $\omega$ which corresponds to a fast exponentially decreasing with $\omega$ term in the exact $K(\omega)$ which should be neglected in the considered region, Eq.(\ref{region}). The term proportional to $g(i\tau)$ after Fourier-transforming gives a tail in $K(\omega)$ in the  region, Eq.(\ref{region}), which is proportional to the power-law spacing distribution $P(s)$:
\begin{equation}\label{Kom_tail}
K(\omega)=\frac{2N^{D_{2}}}{3}\,P(\omega)\sim \frac{N^{-(D_{s}-D_{2})}}{\omega^{1+D_{s}}}.
\end{equation}

\begin{figure}[tbh]
\center{
\includegraphics[width=0.45\linewidth]{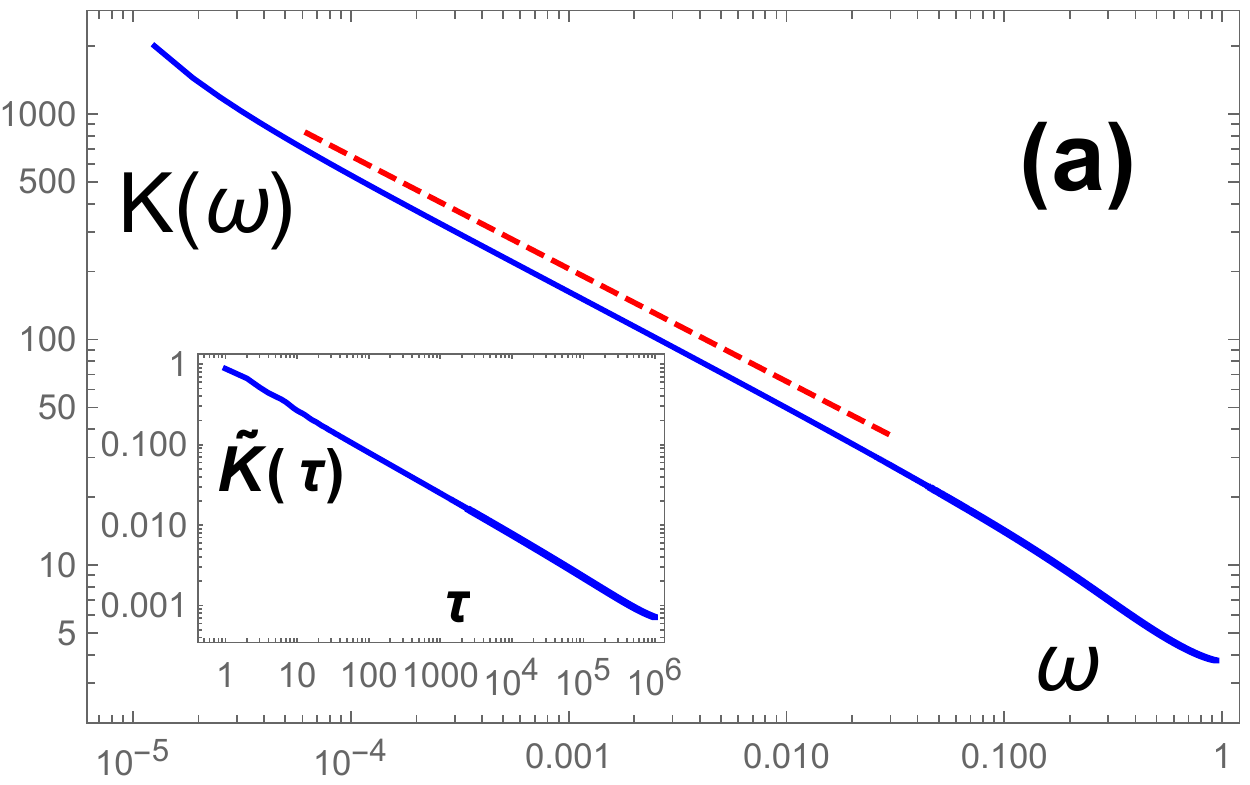}
\includegraphics[width=0.45\linewidth]{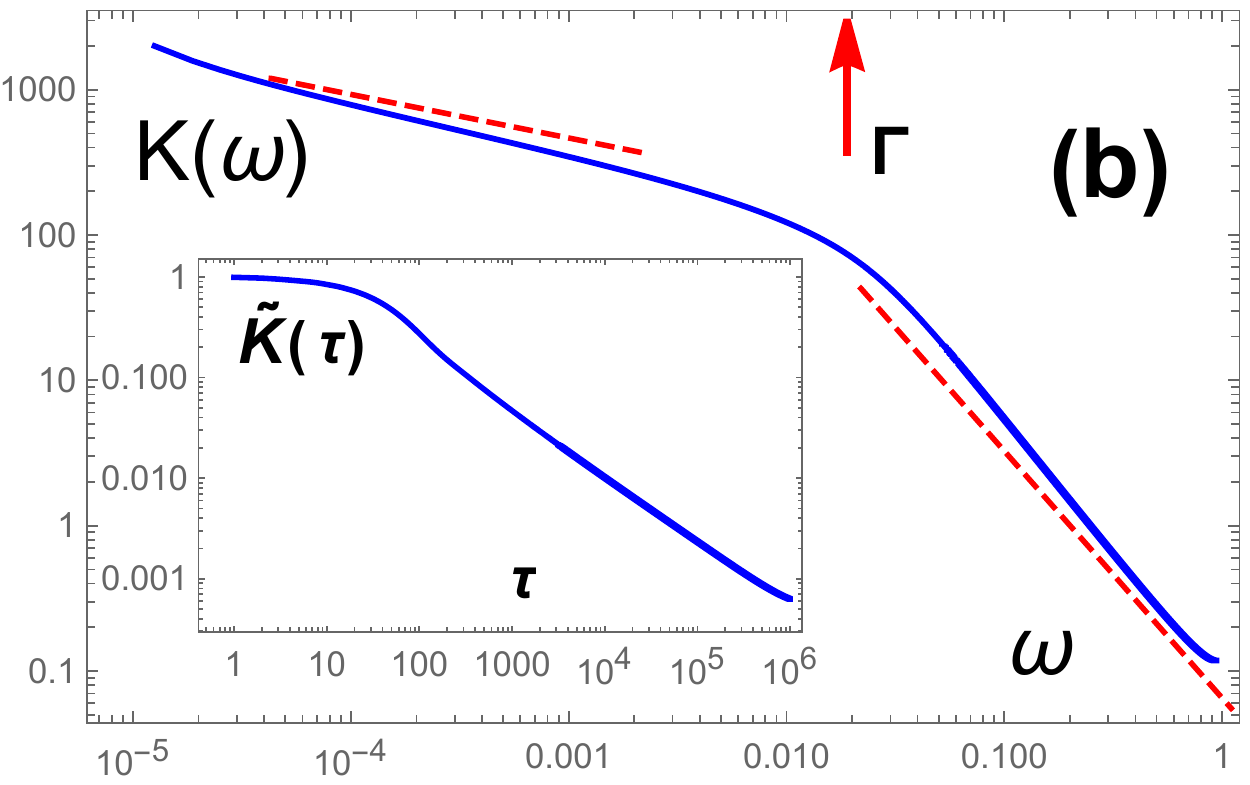}
\includegraphics[width=0.45\linewidth]{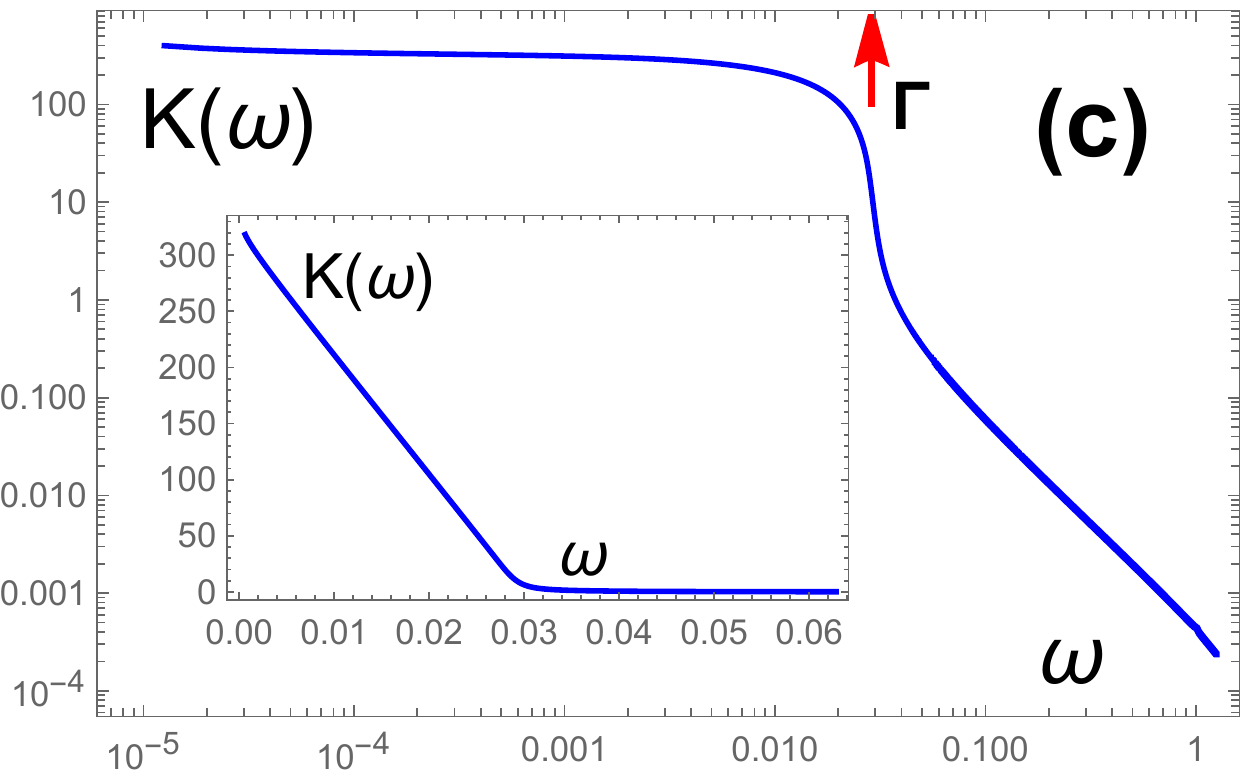}}
\caption{(Color online) The model LDoS correlation function computed exactly from Eqs.(\ref{P_s}),(\ref{K-q}) for: $(a)$ a random Cantor set with $D=D_{s}=0.5$, $N=10^{6}$; $(b)$ a mini-band with the Cantor set inside, $D=0.5$, $D_{s}=0.7$, $N=10^{6}$; a mini-band with the dense spectrum inside, $D=2/3$, $D_{s}=1.5$, $N=10^{6}$. The red dashed lines represent the power-laws, Eqs.(\ref{K_head}),(\ref{Kom_tail}), expected at $N\rightarrow\infty$. Insets: $(a)$ a power-law survival probability  $\tilde{K}(\tau)$; $(b)$ a slow dynamics of survival probability $\tilde{K}(\tau)$ at $\tau<1/\Gamma$ and a power-law tail; $(c)$ a quasi-linear behavior of $K(\omega)$ for $\delta<\omega<\Gamma$.    }
\label{Fig:Kom_exact}
\end{figure}

For $\omega$ smaller than the lower bound of Eq.(\ref{region})
the boundary term $g^{M}$ can be neglected and the results, Eqs.(\ref{K-omega}),(\ref{K_D_s_1}),(\ref{K_om_erg}), of the previous section hold, if one sets $M\sim N^{D_{2}}$:
\begin{equation}\label{K_head}
K(\omega)=\left\{\matrix{N^{D_{s}-D_{2}}\,\omega^{-(1-D_{s})}, & $if$\;\;\;D_{2}<D_{s}<1\cr
N^{1-D_{2}}\,\ln^{-1}(\omega/\delta), & $if$\;\;\;D_{2}<D_{s}=1\cr
N^{1-D_{2}},& $if$\;\;\;D_{2}\leq 1, D_{s}>1 } \right.
\end{equation}
The function $K(\omega)$ at different relationship between $D_{2}\equiv D$ and $D_{s}$ is plotted in Fig.\ref{Fig:Kom_exact}.

Eqs. Eq.(\ref{Kom_tail}) and (\ref{K_head}) show that in order to reproduce the known Lorenzian $K(\omega)$ in the Rosenzweig-Porter RMT one should set $D_{s}=1+\epsilon\equiv 1_{+}$, where $\epsilon\rightarrow +0$:
\begin{equation}\label{K-D_s_1}
K(\omega)\sim \left\{\matrix{ N^{1-D_{2}}, & \delta\ll\omega\ll\Gamma= N^{-(1-D_{2})}\cr
\frac{N^{-(1-D_{2})}}{\omega^{2}},& \omega\gg \Gamma=N^{-(1-D_{2})}}\right..
\end{equation}
The constant $K(\omega)$ at $\delta\ll \omega\ll \Gamma$ (see Fig.\ref{Fig:Kom_exact}(c)) which is similar to that for the Wigner-Dyson RMT, signals of the dense spectrum of levels inside a mini-band (see Fig.\ref{Fig:mini-band} and the inset in Fig.\ref{Fig:C_random_Cantor}(d)). According to our classification presented in Introduction, such type of the spectrum is neither $l$-pure point, since the number of levels $M\rightarrow \infty$ in the limit $N\rightarrow\infty$, nor it is $l$-absolutely continuous, as the width of a mini-band $\Gamma\rightarrow 0$ in this limit.  The singularity of such a spectrum is corroborated by the fact that in the limit of an infinite system $N\rightarrow\infty$ the correlation function $K(\omega)$, Eq.(\ref{K-D_s_1}), reduces to a singular $\delta(\omega)$ function.

Remarkably, Eq.(\ref{K-D_s_1}) reproduces both the behavior of $K(\omega)$ for the Rosen-\\zweig-Porter RMT \cite{RP15} in all limiting cases  and the scaling of the width of a mini-band $\Gamma\sim N^{-(1-D_{2})}$ with the system size $N$. Generalizing this result we conclude from the lower bound in Eq.(\ref{region})  that the width of a mini-band (seen in $K(\omega)$) in a generic case $D_{2}<D_{s}$ is given by:
\begin{equation}\label{Gamma}
\Gamma\sim \left\{\matrix{N^{-1+\frac{D_{2}}{D_{s}}}, & $if$\;\;\; D_{s}<1\cr
N^{-1+D_{2}}\ln(N^{D_{2}}), & $if$\;\;\; D_{s}=1\cr
N^{-1+D_{2}}, & $if$\;\;\; D_{s}>1}\right..
\end{equation}
Notice that Eq.(\ref{Kom_tail}) and (\ref{K_head}) do not match at $\omega\sim \Gamma$ if $D_{s}\geq 1$. This happens because of the narrow    interval   around $\omega=\Gamma$ of the width $\sim \Gamma N^{-D(D_{s}-1)}\ll \Gamma$ where a sharp   drop of $K(\omega)$ by a factor of $N^{-D(D_{s}-1)}$ occurs before a power-law tail, Eq.(\ref{Kom_tail}), becomes the leading contribution. At large $N$ the amplitude of the power-law tail tends to zero at any $D_{s}>1$, so that the edge of a mini-band is   sharp on the logarithmic scale.   This region of a sharp drop is encoded in the $\tau$-independent term in Eq.(\ref{K_tail}) and its detailed description is only possible by a full expression, Eq.(\ref{K-q}). In Fig.\ref{Fig:Kom_exact}(c) it is shown that this drop is in fact linear in $\omega$: $K(\omega)\sim N^{1-D}\,(1-\omega/\Gamma)$.
%The shape of the function $K(\omega)$ at $D_{s}>1$ is  similar to the PDF of the Levy $\alpha$-stable %distribution with $\alpha=D_{s}$, where a similar sharp drop happens at $\alpha>1$ and which is exactly the %Lorentzian at $\alpha=1$. However, $K(\omega)$ has an additional parameter, $N^{-D(D_{s}-1)}$ that controls the %depth of the exponential drop, while the PDF of the (symmetirc) Levy $\alpha$-stable distribution has only two %parameters: $\Gamma$ and $\alpha=D_{s}$.
%%%%%%%%%%%%%%%%%%%%%%%%%%%%%%%%%%%%%%%%%%%%%%%%%%%%%%%%%%%%%%%%%%%%%%%%%%%%%%%%%%%%%%%%%%%%%%%%%%%%%%%%%%%%%%%%%%%%%%%%%%%%%%%%%%%%%%%%%%%%%
%%%%%%%%%%%%%%%%%%%%%%%%%%%%%%%%%%%%%%%%%%%%%%%%%%%%%%%%%%%%%%%%%%%%%%%%%%%%%%%%%%%%%%%%%%%%%%%%%%%%%%%%%%%%%%%%%%%%%%%%%%%%%%%%%%%%%%%%%%%%%

 Concluding this section we would like to emphasize that in general, the exponent $D$  in the definition of the number of levels in the local spectrum:
\begin{equation}\label{D}
M\equiv N^{D}<N^{D_{s}},\;\;\;\;\;(D<D_{s}),
\end{equation}
may depend on the way the discrimination of states is implemented and on the type of the local measure. In a particular case when the local measure is $K(\omega)$ we have $D=D_{2}$ but for a generic local measure in a
 multifractal phase one may have a different $0<D<1$ which depends on the spectrum of multifractal dimensions $D_{q}$.

If $D<D_{s}$, a special type of the local singular-continuous spectrum, {\it the isolated mini-band}, emerges.
  For $D<D_{s}<1$ the levels inside a mini-band form a random Cantor set with a hierarchical structure of higher-order mini-bands (see Fig.\ref{Fig:C_random_Cantor}(a)), while at $D_{s}>1$
the spectrum inside a mini-band is dense (see Fig.\ref{Fig:C_random_Cantor}(d)). At $D_{s}=1$ a special case of non-hierarchical and not completely dense spectrum (the {\it sub-dense} spectrum) shown in Fig.\ref{Fig:C_random_Cantor}(b) emerges inside a mini-band.

Note that the case $D_{s}<D\leq 1$ in Eq.(\ref{P_s}) is inconsistent with  the assumption that the typical level spacing $\delta_{{\rm typ}}$ in the local spectrum is the same as in the global one $\delta_{{\rm typ}}\sim \delta$. Indeed, with this assumption the total local spectral band-width is given by Eq.(\ref{Gamma}): $W_{{\rm local}}\sim N^{\frac{D}{D_{s}}-1}  $. By the discrimination procedure that eliminates levels from the global spectrum in order to obtain a local one, it must be equal or smaller than the total global spectral band-width $W_{{\rm glob}}=1$. This leaves only two options $D=D_{s}$ (random Cantor set) or $D<D_{s}$ (isolated mini-band). For the same reason the case $D=D_{s}=1$ is marginally inconsistent with the above assumption, since in this case $W_{{\rm loc}}\sim \ln(N)>1$.

%%%%%%%%%%%%%%%%%%%%%%%%%%%%%%%%%%%%%%%%%%%%%%%%%%%%%%%%%%%%%%%%%%%%%%%%%%%%%%%%%%%%%%%%%%%%%%%%%%%%%%%%%%%%%%%%%%%
\section{Hierarchy of mini-bands}
%%%%%%%%%%%%%%%%%%%%%%%%%%%%%%%%%%%%%%%%%%%%%%%%%%%%%%%%%%%%%%%%%%%%%%%%%%%%%%%%%%%%%%%%%%%%%%%%%%%%%%%%%%%%%%%%%%%
 \begin{figure}[t]
\center{
\includegraphics[width=0.8\linewidth]{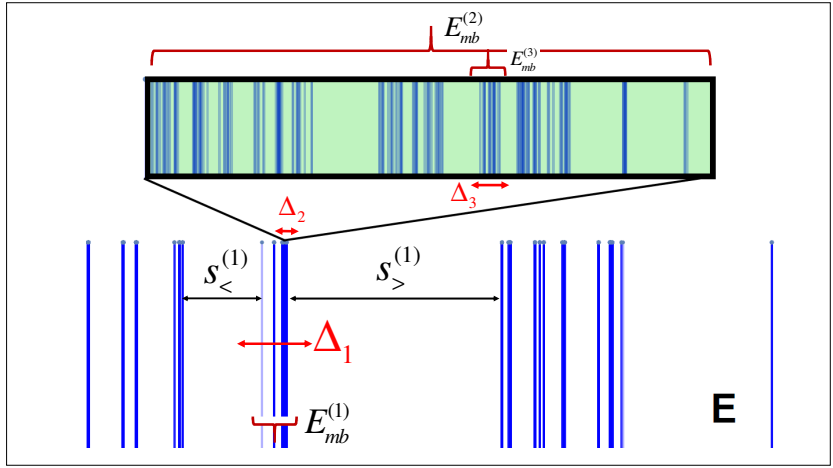}
}
\caption{(Color online) A hierarchy of mini-bands. At each scale $\Delta_{i}$ $(i=1,2,3...)$ one can identify a mini-band of the width $E_{mb}^{(i)}$ such that $E_{mb}^{(i)}\leq \Delta_{i}$ and the spacings $s_{>}^{(i)}, s_{<}^{(i)}$ on the right and on the left of the mini-band obey ${\rm min}\{s_{>}^{(i)},s_{<}^{(i)}\}\geq \Delta_{i}$. A mini-band that at a scale $\Delta_{1}$ looks like a single level at a scale $\Delta_{2}$ (after a zoom) appears to be a mini-band containing  yet a narrower  mini-band at a scale $\Delta_{3}$. }
\label{Fig:hierarchy_minibands}
\end{figure}
%%%%%%%%%%%%%%%%%%%%%%%%%%%%%%%%%%%%%%%%%%%%%%%%%%%%%%%%%%%%%%%%%%%%%%%%%%%%%%%%%%%%%%%%%%%%%%%%%%%%%%%%%%%%%%%%%%%
As it is shown in Fig.\ref{Fig:reg_Cantor} and Fig.\ref{Fig:C_random_Cantor}(a) the regular and a random Cantor set can be considered  as a {\it hierarchy of mini-bands}.

By a hierarchy of mini-bands  we understand a spectrum such that at each of progressively decreasing scales $\Delta_{i}$ ($i=1,2.3...$) one finds a mini-band, i.e a set of consecutive levels obeying the following conditions (see Fig.\ref{Fig:hierarchy_minibands}):
\begin{itemize}
\item
{\it all} the spacings between them are smaller than $\Delta_{i}$
\item their total sum, the width of a mini-band $E_{mb}^{(i)}$, obeys $E_{mb}^{(i)}\leq\Delta_{i}$,
\item
the spacings on the right and on the left of a mini-band $s_{>}^{(i)}$ and $s_{<}^{(i)}$ obey the inequality ${\rm min}\{s_{>}^{(i)},s_{<}^{(i)} \}\geq \Delta_{i}$.
\end{itemize}
Indeed, one can easily see that the regular Cantor set is such a hierarchy of mini-bands at {\it any} legitimate scale $\Delta_{m}=3^{-m}$ bounded only by the minimal scale $\Delta=\delta$ from below and by the total spectral band-width 1 from above. Indeed, at the scale $\Delta_{1}=1/3$ the width  of the two mini-bands $[0,1/3]$ and $[2/3,1]$ with the maximal spacing inside them $1/9 <\Delta_{1}$, are equal to $\Delta_{1}$; at the next scale $\Delta_{2}=1/9$ one has 4 mini-bands of the width $\Delta_{2}=1/9$ with the maximal spacing $1/27<\Delta_{2}$ and so on. Notice that the width of a mini-band at each scale $\Delta_{i}$ is equal to $\Delta_{i}$.
The same property (by the order of magnitude) has a random Cantor set (see Fig.\ref{Fig:C_random_Cantor}(a) and corresponding figure captions).

 %%%%%%%%%%%%%%%%%%%%%%%%%%%%%%%%%%%%%%%%%%%%%%%%%%%%%%%%%%%%%%%%%%%%%%%%%%%%%%%%%%%%%%%%%%%%%%%%%%%%%%%%%%%%%%%%
 \begin{figure}[t]
\center{\includegraphics[width=0.45\linewidth]{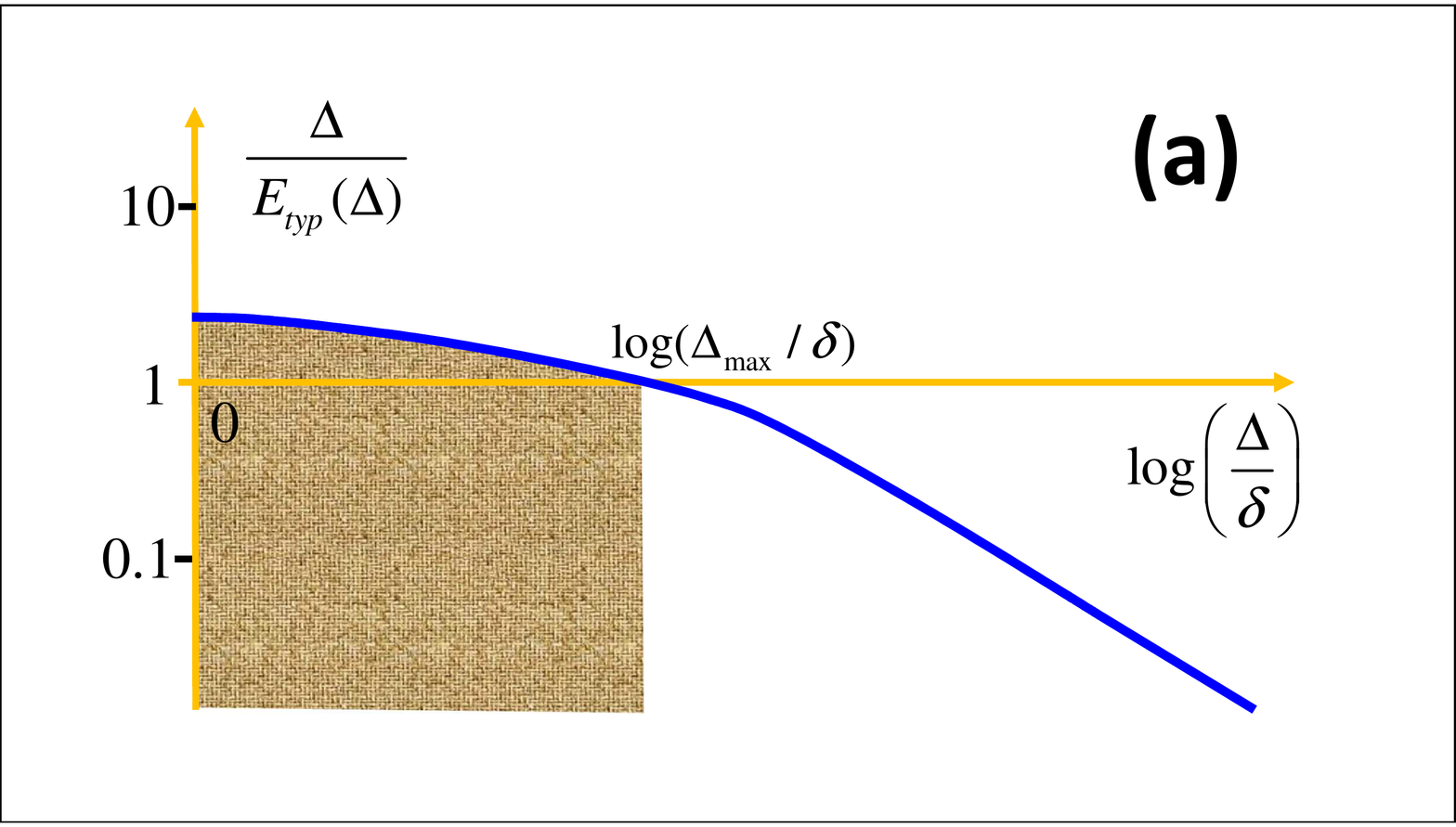}
\includegraphics[width=0.45\linewidth]{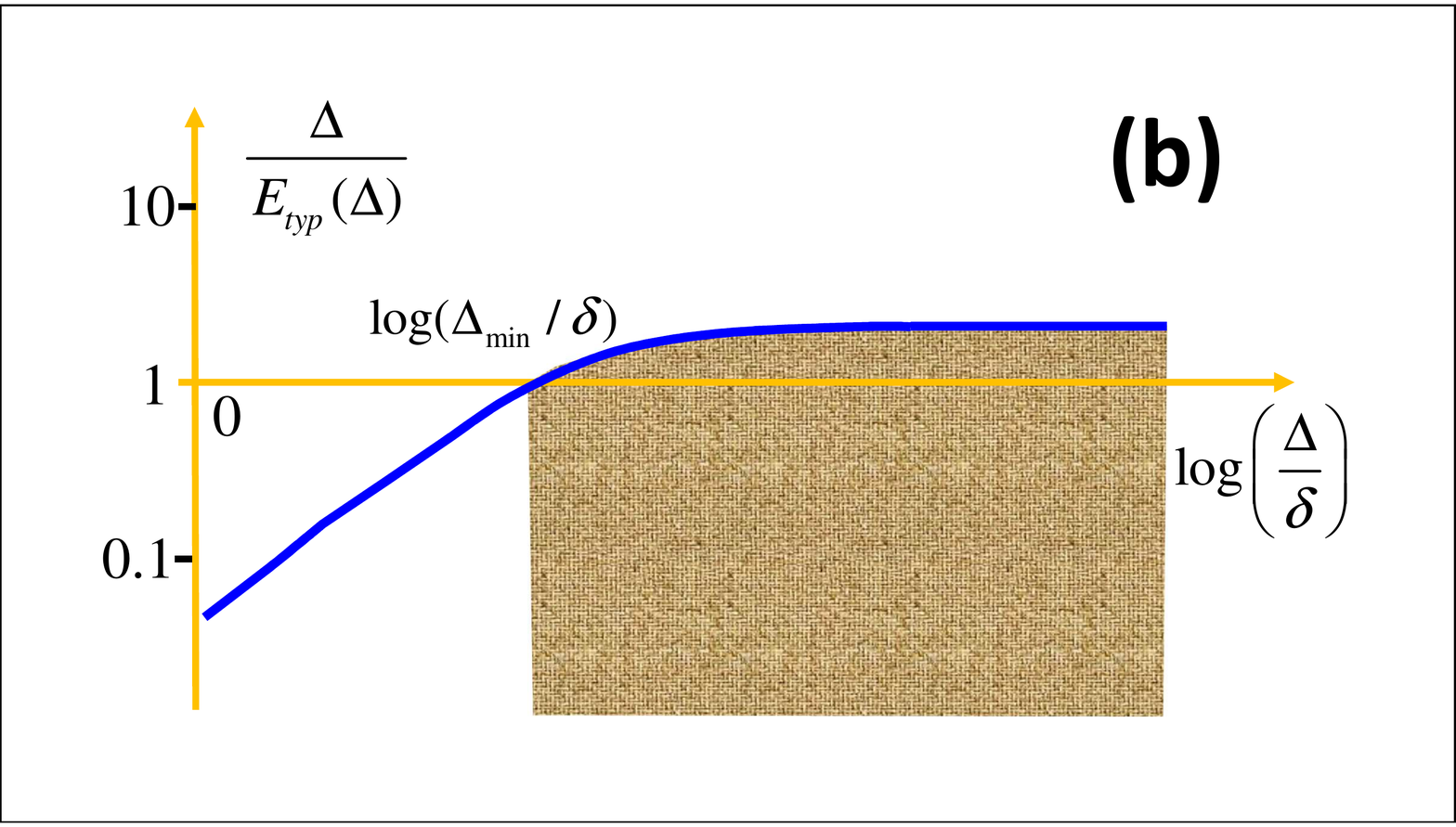}
\includegraphics[width=0.45\linewidth]{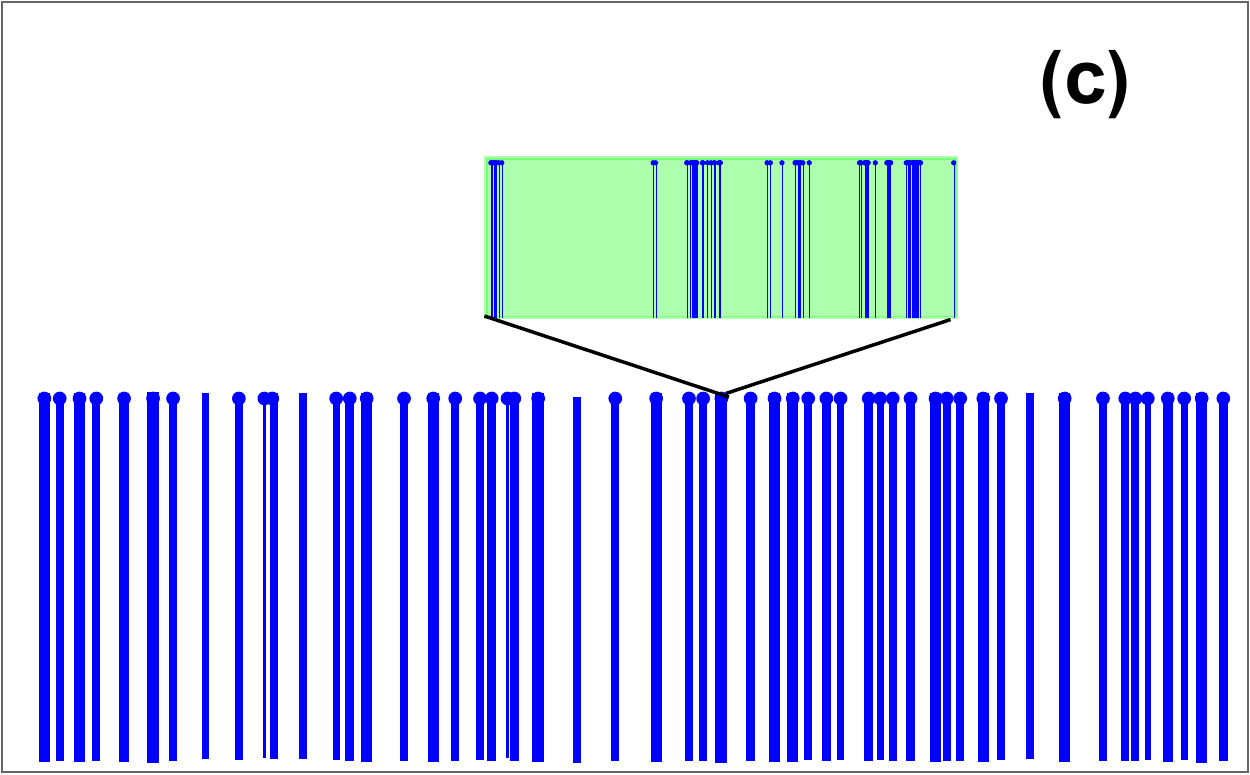}
\includegraphics[width=0.45\linewidth]{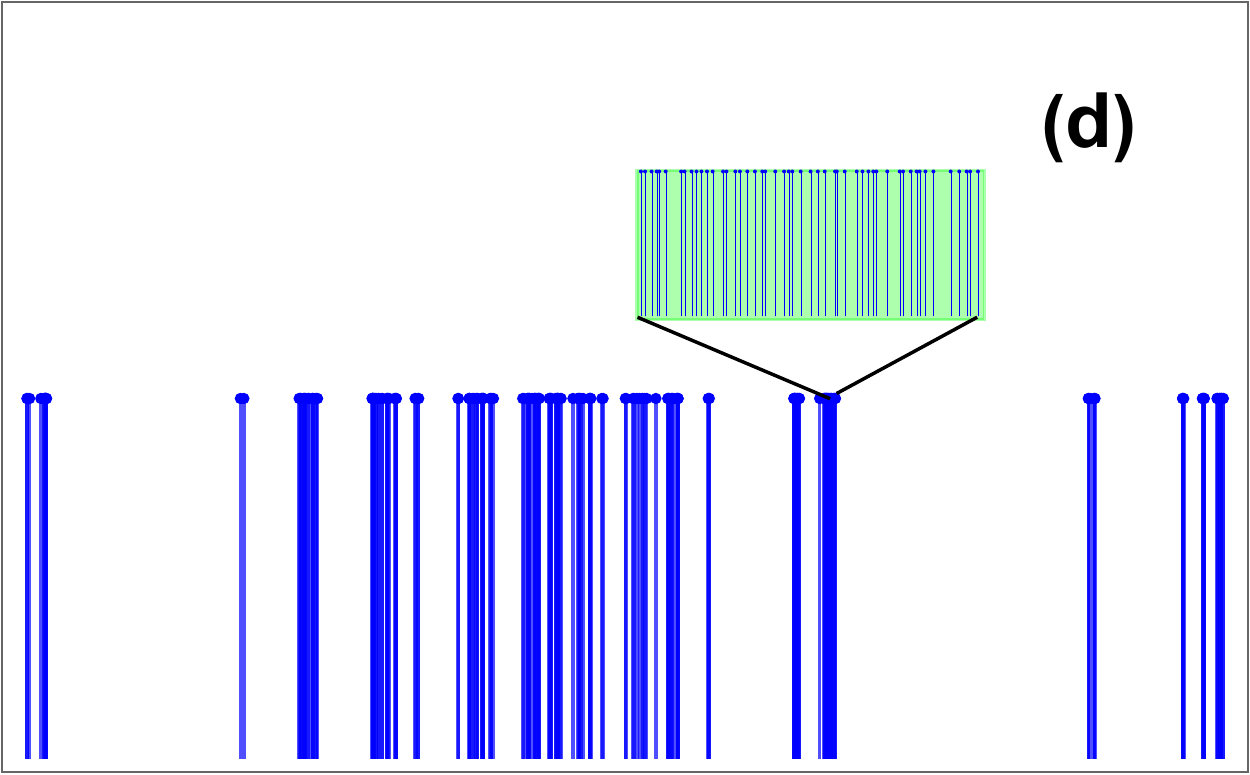}}
\caption{(Color online)   More complicated local spectra. Fractality at small scales $\Delta<\Delta_{{\rm max}}$ and a non-hierarchical set of mini-bands at large scales $\Delta>\Delta_{{\rm max}}$ $[(a),(c)]$. A dense spectrum at small scales $\Delta<\Delta_{{\rm min}}$ and a hierarchical set of mini-bands at large scales $\Delta>\Delta_{{\rm min}}$ $[(b),(d)]$. The region of a hierarchical spectrum in $(a)$ and $(b)$ is shown by a texture.  }
\label{Fig:inner-outer}
\end{figure}
%%%%%%%%%%%%%%%%%%%%%%%%%%%%%%%%%%%%%%%%%%%%%%%%%%%%%%%%%%%%%%%%%%%%%%%%%%%%%%%%%%%%%%%%%%%%%%%%%%%%%%%%%%%%%%
However, the hierarchy of mini-bands is not necessarily extended throughout the entire local spectrum.
 It may terminate from below prior to reaching $\Delta=\delta$. This happens if below some scale $\Delta_{{\rm min}}$ the local spectrum is dense, i.e. $\delta_{{\rm av}}\sim \delta_{{\rm typ}}\sim \delta$ (see Fig.\ref{Fig:inner-outer}(b),(d)). It can also terminate from above at $ \Delta_{\rm max}<1$  if a mini-band of the width $\Gamma\sim \Delta_{{\rm max}}$ is isolated or if the gaps  greater than $\Delta_{{\rm max}}$ {\it between the mini-bands}   are   all of the same order (see Fig.\ref{Fig:inner-outer}((a),(c)). In particular, an isolated mini-band of Fig.\ref{Fig:mini-band} with the dense spectrum inside realized in the Rosenzweig-Porter RMT, does not meet the above criteria of a hierarchy at any scale $\Delta$, as in this case $\Delta_{{\rm min}}=\Delta_{{\rm max}}=\Gamma$.

In order to be able to describe such more complicated than a simple Cantor set cases we present the above conditions for the hierarchy of mini-bands in a form of a mathematical expression in terms of an {\it arbitrary} i.i.d. spacing distribution $P(s)$.

%%%%%%%%%%%%%%%%%%%%%%%%%%%%%%%%%%%%%%%%%%%%%%%%%%%%%%%%%%%%%%%%%%%%%%%%%%%%%%%%%%%%%%%%%%%%%%%%%%%%%%%%%%%%%%%%%%
\subsection{Criterion for existence of a hierarchy of mini-bands}
%%%%%%%%%%%%%%%%%%%%%%%%%%%%%%%%%%%%%%%%%%%%%%%%%%%%%%%%%%%%%%%%%%%%%%%%%%%%%%%%%%%%%%%%%%%%%%%%%%%%%%%%%%%%%%%%%%%
In this section we derive a quantitative criterion of existence of a hierarchy of mini-bands and compute $\Delta_{{\rm min}}$ and $\Delta_{{\rm max}}$ for an arbitrary distribution $P(s)$ of independent identically distributed local spacings.
We start by the expression for the probability for a string of $n$ consecutive levels to obey the following conditions:
\begin{itemize}
\item
  all $(n-1)$ independently fluctuating spacings between them are smaller than some $\Delta$,
\item
 the spacing $s_{>}$ between  the last level in a string  and its nearest neighbor on the right  is greater or equal than $\Delta$.
\footnote{There is no need to impose the condition that the spacing $s_{<}$ between the first level in a string and its nearest neighbor on the left is greater or equal than $\Delta$. This condition is automatically fulfilled, as the neighboring string on the left must have a gap $s_{>}'=s_{<}$ which is greater than $\Delta$. }
\end{itemize}
This probability is given by:
\begin{equation}\label{p-n}
p_{\Delta}(n)=Q\,(1-Q)^{n-1}\approx Q\,e^{-n\,Q},
\end{equation}
where $Q\ll 1$ is the probability for a {\it given} spacing to be greater than $\Delta$:
\begin{equation}\label{Q}
Q\equiv Q(\Delta)=\int_{\Delta}^{1} P(s)\,ds.
\end{equation}
Now let us define the conditional probability  ${\cal P}(E|n)$ that the string of $n$ levels has a width $E$:
\begin{equation}\label{calP_E}
{\cal P}(E|n)=\prod_{k=1}^{n-1}\int ds_{k}\,P(s_{k})\,\delta\left(\sum_{i=1}^{n-1}s_{i}-E\right).
\end{equation}
For independently fluctuating spacings $s_{k}$ this probability distribution function (PDF) is easy to calculate in terms of its (simple) Laplace transform:
\begin{equation}\label{calP_t}
\tilde{{\cal P}}(t|n)=\int_{0}^{\infty}{\cal P}(E|n)\,e^{-t\,E}=[g(t)]^{n-1},
\end{equation}
where $g(t)$ is defined by Eq.(\ref{q}).

The PDF $P_{\Delta}(E)$ of the width $E$ of a string of $n$ levels with all spacings $s_{i}<\Delta$ separated from the rest of the spectrum by two  gaps of the size larger than (or equal to) $\Delta$ is then given by:
\begin{equation}\label{P_Delta_E}
P_{\Delta}(E)=\sum_{n=1}^{M}{\cal P}(E|n)\,p_{\Delta}(n).
\end{equation}
Now we impose the last and the most important condition to be fulfilled for  such a string of $n$ levels to be a mini-band. It is the condition that the {\it typical} width of the string $ E_{{\rm typ}}(\Delta)$ is smaller than or equal to $\Delta$:
\begin{equation}\label{typical_E}
E_{{\rm typ}}(\Delta)={\rm exp}\left[\int \ln(E)\, P_{\Delta}(E)\,dE\right]\lesssim \Delta.
\end{equation}
Eq.(\ref{typical_E}) is one of the principal results of this paper. Together with Eqs.(\ref{p-n})-(\ref{P_Delta_E}) it defines the region where the hierarchy of mini-bands, or the fractality of local spectrum, is present (see Fig.\ref{Fig:inner-outer}).   The boundaries $\Delta_{{\rm min}}$ and $\Delta_{{\rm max}}$ of this region  are given by the  solutions for $\Delta$ of the {\it equation} at an equality sign in Eq.(\ref{typical_E}).
%%%%%%%%%%%%%%%%%%%%%%%%%%%%%%%%%%%%%%%%%%%%%%%%%%%%%%%%%%%%%%%%%%%%%%%%%%%%%%%%%%%%%%%%%%%%%%%%%%%%%%%%%%%%%%%%%%%
\subsection{Examples of a random Cantor set and the dense and sub-dense spectrum}
%%%%%%%%%%%%%%%%%%%%%%%%%%%%%%%%%%%%%%%%%%%%%%%%%%%%%%%%%%%%%%%%%%%%%%%%%%%%%%%%%%%%%%%%%%%%%%%%%%%%%%%%%%%%%%%%%%%
In order to test our approach we compute in Appendix A  the typical width of a mini-band $E_{{\rm typ}}(\Delta)$ for the case of a power-law $P(s)$, Eq.(\ref{P_s}), and show that for $\Delta\gg\delta$:
\begin{eqnarray}
\frac{\Delta}{E_{{\rm typ}}(\Delta)}\sim \left\{\matrix{1, & $for a random Cantor set$\;\; D=D_{s}<1\cr
\ln^{-1}\left(\frac{\Delta}{\delta}\right)\ll 1, & $for a sub-dense spectrum$\;\; D=D_{s}=1 \cr
\left(\delta/\Delta\right)^{D_{s}-1}\ll 1, & $for a dense spectrum$\;\; D=1,\; D_{s}>1} \right.
\end{eqnarray}
As expected, this calculation showed that only the case of a random Cantor set obeys the criterion of fractality, Eq.(\ref{typical_E}), while the dense and sub-dense spectra violate it. More complicated examples are illustrated in Fig.\ref{Fig:inner-outer}

%%%%%%%%%%%%%%%%%%%%%%%%%%%%%%%%%%%%%%%%%%%%%%%%%%%%%%%%%%%%%%%%%%%%%%%%%%%%%%%%%%%%%%%%%%%%%%%%%%%%%%%%%%%%%%%%%
\section{Conclusion}		
%%%%%%%%%%%%%%%%%%%%%%%%%%%%%%%%%%%%%%%%%%%%%%%%%%%%%%%%%%%%%%%%%%%%%%%%%%%%%%%%%%%%%%%%%%%%%%%%%%%%%%%%%%%%%%%%%
The goal of this paper was to explain in simple terms what does the singular-continuous spectrum mean from the physics point of view. In contrast to the mathematical literature where a continuous limit is always assumed, we  consider a discrete system of a finite (but large) number of degrees of freedom $N$  assuming a coarse-graining at small distances (e.g. a lattice with a fixed lattice constant $a=1$) and  taking the   limit $1/N\rightarrow 0$   only in the final results.

Starting from a regular textbook Cantor set we increased complexity of the model   of singular continuous spectrum by considering first a {\it random} Cantor set and then generalizing it to the {\it hierarchy of mini-bands}.

The key ingredient of such constructions was the power-law tailed distribution function $P(s)$ of independent identically distributed spacings between levels seen in the {\it local density of states} (LDoS). We have shown how such a tailed distribution emerges in the limit $N\rightarrow\infty$ as the result of discrimination  that discards states with the amplitude in the observation point smaller than $N^{-1}$. This mechanism explains a qualitative difference between the fractal local spectrum (seen in LDoS) and the dense global spectrum (seen in the global DoS)in the non-ergodic extended phases of disordered systems.

We have demonstrated that within our simple model there are two classes of the local singular-continuous spectra in the non-ergodic extended phases: (i) the random Cantor set and (ii) the isolated mini-band. They are realized in the critical Power-Law Banded Random Matrices (CPLBRM) \cite{PLBRM,KrMuttalib97,CueKrav07,OssKr10} and in the Rosenzweig-Porter (RP) RMT \cite{RP15,FacoetiBiroli_RP,LN-RP20}, respectively. Using the model of independent identically distributed (i.i.d.) local level spacings we reproduced in both cases (i) and (ii) the correlation functions $K(\omega)$ of  the local densities of states  which was previously known from the analytical and numerical studies \cite{KrMuttalib97,CueKrav07,OssKr10,RP15,Param_GRP22, Monthus17a, Monthus17b}. This confirms our conjecture that the simplified picture of i.i.d. local level spacings with the power-law distribution is a correct qualitative description of the above two limiting cases and that the  CPLBRM and RP are characterized by the local singular continuous spectrum of the type of the Cantor set and of the type of the isolated mini-band, respectively.

Our model is characterized\footnote {besides the typical level spacing $\delta=1/N$ which in extended phases is supposed to be of the same order both in the global and in the local spectrum} by two fractal dimensions: the eigenfunction fractal dimension $D$ that describes the number of states $N^{D}$ (with the typical amplitude of wave function $|\psi|^{2}\sim N^{-D}$) seen in the local spectrum after the discrimination, and the spectral fractal dimension $D_{s}$ which determines the power-law distribution of local spacing, Eq.(\ref{P_s}). Depending on the relationship between them, different types of local spectra emerge from our model. All our results are summarized in  Table 1.

\begin{table}
\caption{{\bf  The character of local spectrum, $K(\omega)$ and width of a mini-band $\Gamma$}
}
\label{T:summary}
%\begin{ruledtabular}
 \resizebox{1.0\columnwidth}{!}{%
\begin{tabular}{|c| c| c | c|}
\hline
{\it condition} & {\it local spectrum} & $K(\omega)$ & $\Gamma$ \\
\hline\hline
 $D=D_{s} < 1$ & random Cantor set  & $\omega^{-(1-D_{s})}, \delta<\omega<1$ & 1 \\
\hline
$D=1, D_{s}>1$ & dense set of levels & $\sim 1 $ & 1 \\
\hline
 &  isolated mini-band &  $\Gamma^{-D_{s}}\omega^{-(1-D_{s})}$,  $\delta<\omega<\Gamma$,
 &   \\
  $D<D_{s}<1$ & random Cantor set inside & $\Gamma^{D_{s}}\omega^{-(1+D_{s})}$, $\Gamma<\omega<1$ & $N^{-(1-D/D_{s})}$ \\
\hline
  & isolated mini-band  & $\Gamma^{-1}\ln(N^{D})\ln^{-1}(\omega/\delta)$,  $\delta<\omega<\Gamma$,  &   \\
 $ D<D_{s}=1$& sub-dense spectrum inside   & $\Gamma \omega^{-2}\,\ln^{-1}(N^{D})$,  $\Gamma<\omega<1$ & $N^{-(1-D)}\ln(N^{D})$  \\
  \hline
	&isolated mini-band & $\Gamma^{-1}$,  $\delta<\omega<\Gamma$   &  \\
	$D<1,D_{s}>1$ & dense spectrum inside & $\Gamma^{-1}(\Gamma/\omega)^{1+D_{s}}\,N^{-D(D_{s}-1)}$,  $\Gamma<\omega<1$ & $N^{-(1-D)}$\\
	\hline
	$D=D_{s}=1$& inconsistent with & & \\
	$D_{s}<D\leq 1$   &     $\delta_{{\rm typ}}\sim \delta$ & & \\
	\hline
\end{tabular}%
}
%\end{ruledtabular}
\end{table}
Three of these cases shown in the rows $1$ [Fig.\ref{Fig:C_random_Cantor}(a)], $2$ [Fig.\ref{Fig:C_random_Cantor}(c)] and $5$ [Fig.\ref{Fig:C_random_Cantor}(d)]  have  well-known Hamiltonian realizations: the multifractal states of the critical Power-Law Banded Random Matrices and those at the critical points of the Anderson localization transitions (row 1), the ergodic states of the Wigner-Dyson RMT (row 2) and the  fractal states in the Gaussian Rosenzweig-Porter RMT (row  5  with $D_{s}=1+\epsilon$, $\epsilon\rightarrow 0$).  We are confident that the cases presented in  row $3$ [a mini-band with a random Cantor set, Fig.\ref{Fig:C_random_Cantor}(a), inside ] and row $4$ [a mini-band with a sub-dense spectrum  of Fig.\ref{Fig:C_random_Cantor}(b) inside] also have their Hamiltonian realizations. To find them is an interesting problem of both fundamental and applied significance \footnote{The two cases presented in the row $6$ are inconsistent with the assumption that the typical level spacing in the local spectrum $\delta_{{\rm typ}}\sim \delta$ is of the same order as that in the global one (see footnote $^{1}$). }.

It is important that in all the cases when the eigenfunction fractal dimension $D<1$ (non-ergodic extended states) our model gives a singular continuous spectrum of the type of a random Cantor set or an isolated mini-band. An absolutely continuous local spectrum of Fig.\ref{Fig:C_random_Cantor}(a) emerges only if $D=1$.

In the limit $N\rightarrow\infty$ we obtain from Table 1:
\begin{equation}\label{Kom_N_inf_lim}
\lim_{N\rightarrow\infty} K(\omega)\sim \left\{\matrix{\omega^{-1+D_{s}}, & $for a random Cantor set$\cr
\delta(\omega), & $for an isolated mini band$ \cr
1, & $for an absolutely continuous spectrum $ } \right.
\end{equation}

Eq.(\ref{Kom_N_inf_lim}) expresses a very general important result: The LDoS correlation function in the limit $N\rightarrow\infty$ is a measure that distinguishes between the absolutely continuous and the singular continuous spectra in delocalized phases. While the former results in a finite limit for $\lim_{N\rightarrow\infty} K(\omega)$ at $\omega\rightarrow 0$, the latter implies the singularity at $\omega=0$. Furthermore, the type of the singularity unambiguously identifies the type of a singular-continuous spectrum. For the localized eigenstates the typical level spacing in the local spectrum is finite in $N\rightarrow \infty$ limit and, hence, $K(\omega\rightarrow 0)\rightarrow 0$ in this limit.

Extending the simple picture described above to an arbitrary distribution of local level spacing, we formulated a general criterion, Eq.(\ref{typical_E}), of fractality of local spectra and tested it by considering the examples of the random Cantor set, the isolated mini-band and the dense spectrum. The criterion Eq.(\ref{typical_E}) is one of the main results of the paper which can be used to characterize more complicated local spectra in the non-ergodic extended phases.

{\bf Acknowledgment.} We are grateful to E. Bogomolny, M. V. Feigel'man, I. M. Khaymovich, P. Nosov, A. Scardicchio
and M. Zirnbauer for stimulating discussions and useful comments on the manuscript.  V.E.K. also appreciates collaboration with L. B. Ioffe and M. Pino during the work on Ref.\cite{Pino-Ioffe-VEK} where  some preliminary results of this paper were communicated. The support
from   the Google
Quantum Research Award “Ergodicity Breaking in Quantum
Many-Body Systems” (V.E.K.) is gratefully acknowledged.
%%%%%%%%%%%%%%%%%%%%%%%%%%%%%%%%%%%%%%%%%%%%%%%%%%%%%%%%%%%%%%%%%%%%%%%%%%%%%%%%%%%%%%%%%%%%%%%%%%%%%%%%%%%%%%%%%%
%%%%%%%%%%%%%%%%%%%%%%%%%%%%%%%%%%%%%%%%%%%%%%%%%%%%%%%%%%%%%%%%%%%%%%%%%%%%%%%%%%%%%%%%%%%%%%%%%%%%%%%%%%%%%%%%%%
\appendix
%%%%%%%%%%%%%%%%%%%%%%%%%%%%%%%%%%%%%%%%%%%%%%%%%%%%%%%%%%%%%%%%%%%%%%%%%%%%%%%%%%%%%%%%%%%%%%%%%%%%%%%%%%%%%%%%%%
\section{A test of Eq.(\ref{typical_E}) for a power-law $P(s)$ in Eq.(\ref{P_s})}
%%%%%%%%%%%%%%%%%%%%%%%%%%%%%%%%%%%%%%%%%%%%%%%%%%%%%%%%%%%%%%%%%%%%%%%%%%%%%%%%%%%%%%%%%%%%%%%%%%%%%%%%%%%%%%%%%%
\subsection{Example of a random Cantor set}
%%%%%%%%%%%%%%%%%%%%%%%%%%%%%%%%%%%%%%%%%%%%%%%%%%%%%%%%%%%%%%%%%%%%%%%%%%%%%%%%%%%%%%%%%%%%%%%%%%%%%%%%%%%%%%%%%%
In order to illustrate a general formalism Eqs.(\ref{p-n})-(\ref{typical_E}) we consider an example of a random Cantor set with $D=D_{s}<1$, where $D$ is defined in Eq.(\ref{D}).

Consider Eq.(\ref{calP_t}) in the limit $n\gg 1$ with $g(t)$ from Eq.(\ref{q-expan}):
\begin{equation}
[g(t)]^{n-1}\approx {\rm exp}[-n (t\,\delta)^{D_{s}}].
\end{equation}
Then the PDF ${\cal P}(E|n)$ is given by the Laplace transform:
\begin{equation}\label{inverse_laplace}
{\cal P}(E|n)=\int_{B} {\rm exp}\left[-n(t\,\delta)^{D_{s}}+E t \right]\,dt,
\end{equation}
where the integral runs over the Bromwich contour $t\in (-i\infty +0, i\infty +0)$ as customary for the inverse Laplace transform \footnote{As a matter of fact, Eq.(\ref{inverse_laplace}) defines a PDF of the {\it Levy $\alpha$-stable distribution}  with $\alpha=D_{s}$. }.

Next we do summation over $n$ according to Eqs.(\ref{P_Delta_E}),(\ref{p-n}):
\begin{equation}\label{P_delta_Bromv}
P_{\Delta}(E)=Q\int_{B} \frac{e^{E t}}{Q+(t\,\delta)^{D_{s}}}\,\frac{dt}{2\pi i},
\end{equation}
 using the smallness of $Q$ and $t\delta$ to approximate $1-e^{-(Q+(t\delta)^{D_{s}})}\approx Q+(t\delta)^{D_{s}}$ and large values of $M (Q+(t\delta)^{D_{s}})\gg 1$ to extend summation over $n$ to infinity.
Integrating $e^{E t}$ over the Bromvich contour first to obtain $\delta(t)$, one can easily check that the distribution function $P_{\Delta}(E)$ is correctly normalized.

The Bromvich contour in Eq.(\ref{P_delta_Bromv}) can be deformed to go along the branch cut $(-\infty-i0, 0)\cup(0,-\infty+0)$. Then switching $t\rightarrow -t$ we obtain:
\begin{equation}
P_{\Delta}(E)=\frac{\sin\left(  \pi D_{s} \right)}{\pi}\,
Q\int_{0}^{\infty} dt\,e^{-E t}\,\frac{(t\delta)^{D_{s}}}{Q^{2}+(t\delta)^{2D_{s}}+2 Q (t\delta)^{D_{s}} \,\cos(\pi D_{s})}.
\end{equation}
The integral in this expression  is controlled by a parameter $E/E_{*}$, where:
\begin{equation}\label{E_star}
E_{*}=\frac{\delta}{Q^{\frac{1}{D_{s}}}},
\end{equation}
and in the limiting cases $E\ll E_{*}$ and $E\gg E_{*}$ it is equal to:
\begin{equation}\label{P_Delta_appr}
P_{\Delta}(E)=\frac{\sin\left( \frac{\pi D_{s}}{2}\right)}{\pi}
\,\left\{\matrix{ \frac{\Gamma(1-D_{s})}{E}
 \,\left(\frac{E}{E_{*}}\right)^{D_{s}}, & (E\ll E_{*})\cr
\frac{\Gamma(1+D_{s})}{E}\,\left(\frac{E_{*}}{E}\right)^{D_{s}}, & (E\gg E_{*})}  \right.
\end{equation}
\begin{figure}[t]
\center{
\includegraphics[width=0.8\linewidth]{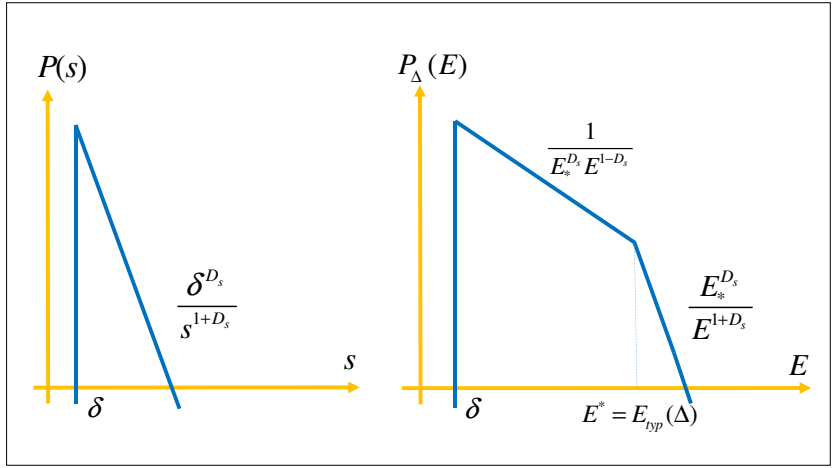}
}
\caption{(Color online) The distribution of local spacings in the log-log coordinates (Left panel) and the corresponding distribution of the width $E$ of a mini-gap at a scale $\Delta$ (Right panel) for the random Cantor set. The main contribution to the normalization integral of $P_{\Delta}(E)$ is given by $E_{*}=E_{{\rm typ}}(\Delta)$.}
\label{Fig:P_s-P_E}
\end{figure}
The sketch of the distribution function $P_{\Delta}(E)$ and the corresponding spacing distribution is shown in Fig.\ref{Fig:P_s-P_E}.
As the integral of $P_{\Delta}(E)$ over $E<E_{*}$ (the normalization integral) is dominated by the upper limit $E\sim E_{*}$, and the corresponding integral over $E>E_{*}$ is dominated by the lower limit $E\sim E_{*}$,  the main contribution to the normalization integral of $P_{\Delta}(E)$ comes from the vicinity of $E\approx E_{*}$ given by Eq.(\ref{E_star}). By definition $E_{*}$ is the typical width $E_{{\rm typ}}(\Delta)$ of a mini-band at a scale $\Delta$.

The last step is to substitute $Q=Q(\Delta)$ in Eq.(\ref{E_star}). For the local spacing distribution function $P(s)$, Eq.(\ref{P_s}), one readily finds:
\begin{equation}\label{Q-Q}
Q(\Delta)\sim \left(\frac{\delta}{\Delta} \right)^{D_{s}},
\end{equation}
so that
\begin{equation}
E_{*}=E_{{\rm typ}}(\Delta)\sim \Delta.
\end{equation}
We conclude, therefore, that for the random Cantor set, like for its regular counterpart, the typical width of a mini-band at a scale $\Delta$ is of the order of $\Delta$. This saturates the criterion, Eq.(\ref{typical_E}) which is therefore fulfilled for {\it any} $\Delta>\delta$. This result was expected but it serves as an important test for the consistency of the formalism Eqs.(\ref{p-n})-(\ref{typical_E}) developed in  Section 7.
Since the random Cantor set is considered as an archetypical example of a fractal spectrum, the fulfillment of the criterion Eq.(\ref{typical_E}) in general signals on the fractality of the local spectrum.
%%%%%%%%%%%%%%%%%%%%%%%%%%%%%%%%%%%%%%%%%%%%%%%%%%%%%%%%%%%%%%%%%%%%%%%%%%%%%%%%%%%%%%%%%%%%%%%%%%%%%%%%%%%%%%%%%%%
\subsection{Example of an isolated mini-band}
%%%%%%%%%%%%%%%%%%%%%%%%%%%%%%%%%%%%%%%%%%%%%%%%%%%%%%%%%%%%%%%%%%%%%%%%%%%%%%%%%%%%%%%%%%%%%%%%%%%%%%%%%%%%%%%%%%%
Now we consider the case   $D<D_{s}$  in Eqs.(\ref{D}),(\ref{P_s}) which corresponds to an isolated mini-band. First of all we note that Eq.(\ref{P_delta_Bromv}) is only valid if one can neglect the boundary term in the summation of geometric series. This necessarily requires $M Q=N^{D} Q\gg 1$, which by the virtue of Eq.(\ref{Q-Q}) implies:
\begin{equation}\label{Delta_max}
\Delta \ll \Delta_{{\rm max}}=\Gamma\sim \left\{ \matrix{N^{-1+\frac{D}{D_{s}}}, & D_{s}<1\cr
N^{-1+D}, & D_{s}\geq 1       } \right..
\end{equation}
Within this restriction let us consider the character of spectrum {\it inside a mini-band} of the width $\Gamma$. If $D_{s}<1$ the results of the previous subsection [Eq.(\ref{K-omega}) with $M\sim N^{D}$] show that the spectrum inside a mini-band  is a random Cantor set with the hierarchy of mini-bands.

If $D<1$ and $D_{s}=1$ we have using Eq.(\ref{q_D_s_1}):
\begin{equation}
P_{\Delta}(E)\approx
\frac{Q}{2}\int_{0}^{\infty} dt\,e^{-E t}\,\frac{(t\delta) }{(Q-(t\delta)\ln(t\delta))^{2}}
=\left\{\matrix{\frac{Q}{2\delta\,\ln^{2}(E/\delta)}, & \delta\ll E\ll E_{*}\cr
\frac{\delta}{2Q E^{2}}, & \Gamma\gg E\gg E_{*}}\right.,
\end{equation}
where $Q=\delta/\Delta$ and
\begin{equation}
E_{*} =\Delta \ln(\Delta/\delta).
\end{equation}
As in the previous subsection, one can see that the main contribution to the normalization integral for $P_{\Delta}(E)$ is given by $E\approx E_{*}$, so that $E_{{\rm typ}}(\Delta)=E_{*}$. Therefore,
\begin{equation}
E_{{\rm typ}}(\Delta)>\Delta
\end{equation}
 in the entire region $\delta\ll \Delta<\Gamma$.  The failure to fulfill the criterion Eq.(\ref{typical_E})  implies that  for $D<D_{s}=1$ the spectrum inside the mini-band does not have a hierarchical structure.
At the same time it has large gaps and thus is not completely dense. The similar spectrum appears in the entire band for  the case $D=D_{s}=1$ (see Fig.\ref{Fig:C_random_Cantor}(b)).

Finally, let us consider the case $D<1$ but $2>D_{s}>1$. In this case it follows from Eq.(\ref{q_D_s_great_1}) that:
\begin{equation}
 P_{\Delta}(E)\approx Q\int_{B} \frac{e^{E t}}{Q+c_{1}(t\delta)+c_{2}(t\,\delta)^{D_{s}}}\,\frac{dt}{2\pi i},
\end{equation}
where $c_{1}$ and $c_{2}$ are coefficients of order 1.

The Bromwich contour $B$ of integration can be deformed to  encircle the cut $(-\infty, 0)$. Then at $E/\delta\gg 1$ the main contribution comes from $t\delta \ll 1$, so that the term $c_{2}(t\,\delta)^{D_{s}}$ is a correction. The leading $t$-dependence leads to a pole at $c_{1} t\delta=Q$ with the result:
\begin{equation}
P_{\Delta}(E)|_{{\rm pole}}=\frac{c_{1}}{E_{*}}\,e^{-(E/E^{*})}
\end{equation}
where
\begin{equation}\label{E_star_gr_1}
E_{*}=\frac{c_{1}\delta}{Q}.
\end{equation}
This pole contribution is almost constant  for $E\ll E_{*}$ and it is exponentially small at $E\gg E_{*}$. In this latter region one should take into account a correction $c_{2}(t\,\delta)^{D_{s}}$ which takes different values at $t=-x+i 0$ and $t=-x-i 0$, $(x>0)$:
\begin{equation}
P_{\Delta}(E)|_{{\rm corr}}=\frac{c_{2}\,\sin(\pi D_{s})}{\pi}\int_{0}^{\infty} dx\,e^{-E x}\,\frac{(x\delta)^{D_{s}}}{Q}
\end{equation}
Collecting the results at $E\ll E_{*}$ and $E\gg E_{*}$ we finally obtain:
\begin{equation}\label{P_Delta}
 P_{\Delta}(E)\sim \left\{\matrix{\frac{e^{-E/E_{*}}}{E_{*}}, &\delta \ll E\ll E_{*}\ln[c( E_{*}/\delta)^{D_{s}-1}]\cr
\left( \frac{\delta}{E}\right)^{D_{s}+1}\frac{E_{*}}{\delta^{2}} , & \Gamma\gg E\gg E_{*}\ln[ c( E_{*}/\delta)^{D_{s}-1}]}\right.,
\end{equation}
 where $c\sim 1$.

One can see again that the main contribution to the normalization integral comes from  $E\sim E_{*}$. So, from Eqs.(\ref{Q-Q}),(\ref{E_star_gr_1}) we obtain:
\begin{equation}\label{EE_typ}
E_{{\rm typ}}(\Delta)\sim E_{*}\sim \Delta\,\left(\frac{\Delta}{\delta}\right)^{D_{s}-1}\gg \Delta,
\end{equation}
for all $\Delta\gg\delta$. This result remains valid for all $D_{s}>1$.

In this case the length of the string of levels with all spacing between them smaller than $\Delta$ is larger by a polynomially  large parameter $(\Delta/\delta)^{D_{s}-1}$ than the gaps between such strings. This makes large gaps extremely rare which means dense typical local spectrum inside a mini-gap at $D<1$ and that in the entire spectral band for $D=1$ (see Fig.\ref{Fig:C_random_Cantor}(c)).
\subsection{Example of a dense spectrum}
This case is easy to obtain from the results of the previous subsection at $D=1$ and $D_{s}>1$. The only modification needed is to replace $\Delta_{{\rm max}}=\Gamma$ given by Eq.(\ref{Delta_max}) by the entire band-width $\Delta_{{\rm max}}=\Gamma=1$. After this modification Eq.(\ref{P_Delta})  remains valid in the entire region $\delta<E<1$, and Eq.(\ref{EE_typ}) gives:
\begin{equation}
\frac{\Delta}{E_{{\rm typ}}(\Delta)}\sim \left(\frac{\delta}{\Delta}\right)^{D_{s}-1}\ll 1.
\end{equation}
Thus the criterion of  fractality of local spectrum, Eq.(\ref{typical_E}), is strongly violated as it should be for the dense spectrum.
%% The Appendices part is started with the command \appendix;
%% appendix sections are then done as normal sections
%% \appendix

%% \section{}
%% \label{}

%% If you have bibdatabase file and want bibtex to generate the
%% bibitems, please use
%%
%%  \bibliographystyle{elsarticle-harv}
%%  \bibliography{<your bibdatabase>}

%% else use the following coding to input the bibitems directly in the
%% TeX file.

%%%%%%%%%%%%%%%%%%%%%%%%%%%%%%%%%%%%%%%%%%%%%%%%%%%%%%%%%5

\bibliographystyle{unsrt}
\bibliography{Efetov}
%%%%%%%%%%%%%%%%%%%%%%%%%%%%%%%%%%%%%%%%%%%%%%%%%%%%%%%%%5

%\begin{thebibliography}{00}

%% \bibitem[Author(year)]{label}
%% Text of bibliographic item

%\bibitem[ ()]{}

%\end{thebibliography}
\end{document}